\documentclass[twocolumn,pra]{revtex4}

\usepackage{graphicx}
\usepackage{amssymb}
\usepackage{epstopdf}
\usepackage{amsmath}
\usepackage{color}
\usepackage{hyperref}

\usepackage[acronym]{glossaries}
\usepackage{physics}
\makeatletter

\@ifundefined{textcolor}{}
{
 \definecolor{BLACK}{gray}{0}
 \definecolor{WHITE}{gray}{1}
 \definecolor{RED}{rgb}{1,0,0}
 \definecolor{GREEN}{rgb}{0,1,0}
 \definecolor{BLUE}{rgb}{0,0,1}
 \definecolor{CYAN}{cmyk}{1,0,0,0}
 \definecolor{MAGENTA}{cmyk}{0,1,0,0}
 \definecolor{YELLOW}{cmyk}{0,0,1,0}
}

\graphicspath{{Images/}}

\hypersetup{
	pdfpagemode={UseOutlines},
	bookmarksopen,
	pdfstartview={FitH},
	colorlinks,
	linkcolor={black},
	citecolor={blue},
	urlcolor={blue}}

\makeatother

\newacronym{GPE}{GPE}{Gross-Pitaevskii Equation}
\newacronym{TWA}{TWA}{Truncated Wigner Approximation}
\newacronym{MI}{MI}{Mott Insulator}
\newacronym{SF}{SF}{Superfluid}
\newacronym{MQST}{MQST}{Macroscopic Quantum Self-Trapping}
\newacronym[plural=FPs,firstplural=Fixed Points (FPs)]{FP}{FP}{Fixed Point}
\newacronym{UC}{UC}{Uniform Configuration}
\newacronym{SPAC}{SPAC}{Symmetric $\pi$-Aligned Configuration}
\newacronym{SFC}{SFC}{Superfluid Configuration}
\newacronym{DC}{DC}{Delta Configuration}
\newacronym[plural=DPTs,firstplural=Dynamical Phase Transitions (DPTs)]{DPT}{DPT}{Dynamical Phase Transition} 

\begin{document}

\title{Quench-induced dynamical phase transitions and $\pi$-synchronization in the Bose-Hubbard model}

\author{Andrea Pizzi$^{1,2,3}$, Fabrizio Dolcini$^{2}$ and Karyn Le Hur$^{1}$}
\affiliation{$^1$ CPHT, Ecole Polytechnique, CNRS, Universit\' e Paris-Saclay, Route de Saclay, 91128 Palaiseau, France}
\affiliation{$^2$ Dipartimento di Scienza Applicata e Tecnologia, Politecnico di Torino, I-10129 Torino, Italy}
\affiliation{$^3$ Cavendish Laboratory, University of Cambridge, Cambridge CB3 0HE, United Kingdom}


\begin{abstract}
We investigate the non-equilibrium behavior of a fully-connected (or all-to-all coupled) Bose-Hubbard model after a Mott to superfluid quench, in the limit of large boson densities and for an arbitrary number $V$ of lattice sites, with potential relevance in experiments ranging from cold atoms to superconducting qubits. By means of the truncated Wigner approximation, we predict that crossing a critical quench strength the system undergoes a dynamical phase transition between two regimes that are characterized at long times either by an inhomogeneous population of the lattice (i.e.~macroscopical self-trapping) or by the tendency of the mean-field bosonic variables to split into two groups with phase difference $\pi$, that we refer to as $\pi$-synchronization. We show the latter process to be intimately connected to the presence, only for $V \ge 4$, of a manifold of infinitely many fixed points of the dynamical equations. Finally, we show that no fine-tuning of the model parameters is needed for the emergence of such $\pi$-synchronization, that is in fact found to vanish smoothly in presence of an increasing site-dependent disorder, in what we call a synchronization crossover.
\end{abstract}

\maketitle

\section{Introduction}

The theory of interacting many-body quantum systems at equilibrium has advanced remarkably over the past few decades, to account for various quantum phase transitions, i.e. sharp changes of the ground state of an Hamiltonian when its parameters are varied  across some critical values. However, the behavior of such systems is far less understood when it comes to the out-of-equilibrium regime, whose relevance has rapidly grown triggered by significant experimental progress in gases of ultra-cold neutral atoms in optical lattices
\cite{greiner2002quantum, bloch2008quantum, esteve2008squeezing, schneider2008metallic, bakr2010probing, abanin2018ergodicity}, trapped atoms \cite{hofferberth2007non, albiez2005direct, schumm2005matter, gati2006noise}, superconducting qubits \cite{roushan2017spectroscopic, xu2018emulating}. One of the most established protocols to take these systems to the non-equilibrium regime is the quantum quench, consisting of a sudden change of the Hamiltonian of the system from $H_i$ to $H_f$ at time $t = 0$ \cite{sengupta2004quench, kollath2007quench, moeckel2008interaction, chin2010feshbach, schiro2010time, sciolla2010quantum, sciolla2011dynamical, sciolla2013quantum}. Importantly, over the typical experimental timescales these systems are essentially isolated from the environment. In these conditions, one can observe the emergence of two particularly interesting phenomena. The first one is a quantum \gls{DPT}, identified by a sharp change of the dynamical behavior at a critical quench strength \cite{sciolla2010quantum, sciolla2011dynamical, sciolla2013quantum, heyl2013dynamical, eckstein2009thermalization}, whereas the second, sometimes referred to as thermalization of an isolated system \cite{flesch2008probing, rigol2008thermalization, cramer2008exploring, cramer2009exact, rigol2009breakdown, eckstein2009thermalization, biroli2010effect, dziarmaga2010dynamics, kollar2011generalized}, consists in the relaxation of some macroscopic variables to some finite values at long times.

A well-established approach to the study of the dynamics of a large population of interacting bosons on a lattice consists of reducing the Heisenberg equation of motion to the discrete nonlinear \gls{GPE} via a mean-field substitution of the bosonic creation and annihilation operators $a_j^\dagger$ and  $a_j$ ($j = 1, 2, \dots, V$ labeling the lattice site) with the $\mathbb{C}$-numbers $\psi_j$ and $\psi_j^*$
\cite{smerzi1997quantum, mossmann2006semiclassical, gati2007bosonic, graefe2008mean, kolovsky2009bloch, witthaut2017classical, raghavan1999coherent}.
Quantum fluctuations can then be taken into accout within the \gls{TWA}, that at each time $t>0$ considers averages over an ensemble of classical fields $\{\psi_j(t)\}$ obtained as the evolution under the \gls{GPE} of stochastic initial fields $\{\psi_j(0)\}$ \cite{polkovnikov2002nonequilibrium, polkovnikov2003quantum, polkovnikov2010phase}.
Looking at the time evolution of the phases $\{\theta_j\}$ of $\{\psi_j\}$, the system can be regarded as a system of $V$ classical nonlinearly coupled oscillators, making thus natural to wonder about the occurrence of synchronization phenomena. Indeed, Witthaut \textit{et al.} recently demonstrated that a particular class of bosonic models can in this way be recasted to the Kuramoto model for classically coupled nonlinear oscillators, that is a well-known model revealing a synchronization transition driven by the competition between coupling and disorder \cite{witthaut2017classical, kuramoto1975self, strogatz2000kuramoto, acebron2005kuramoto}. Importantly, since the considered systems are (almost) isolated, the emergence of synchronization is not due to any dissipation or external driving, as usually considered. \cite{holmes2012synchronization, lee2013quantum, lee2014entanglement, walter2014quantum, bastidas2015quantum, orth2010dynamics, le2018driven, henriet2016quantum, zhu2015synchronization}.

In this context, the Bose-Hubbard model is paradigmatic, describing a system of bosons on a lattice with site-to-site tunneling and on-site interaction, and exhibiting at equilibrium a quantum phase transition between a \gls{SF} and a \gls{MI} \cite{fisher1989boson, freericks1994phase, rachel2012detecting, boeris2016mott, trotzky2012probing}. Such a model finds various applications in physics \cite{cazalilla2011one} from ultra-cold atoms in optical lattices \cite{greiner2002quantum, jaksch1998cold, zwerger2003mott} to systems of Josephson junctions \cite{glazman1997new, kuzmin2018quantum, weissl2015kerr, roushan2017spectroscopic}. For these systems, disorder is known to lead to glassy phases and Anderson localization \cite{fisher1989boson, ristivojevic2012phase, jendrzejewski2012three, doggen2017weak, giamarchi1987localization}, while recently the phenomenon of many-body localization has also been analyzed \cite{abanin2018ergodicity}. In the non-equilibrium regime, using an exact approach Sciolla and Biroli highlighted the existence of a \gls{DPT} for a fully-connected lattice in the limit of infinite site number ($V \rightarrow \infty$) for small boson densities (that is few bosons per site) \cite{sciolla2010quantum, sciolla2011dynamical, sciolla2013quantum}. In the opposite limit of large boson densities, the \gls{TWA} was adopted to study one, two and three-dimensional systems \cite{polkovnikov2002nonequilibrium, dziarmaga2012quench}, whereas other works focused on the bosonic dimer and trimer (that is $V = 2, 3$) revealing peculiar dynamical features such as the \gls{MQST}, that is a symmetry breaking leading to non-zero average population imbalance \cite{milburn1997quantum, smerzi1997quantum, franzosi2000quantum, gati2007bosonic, graefe2008mean, longhi2011optical, franzosi203509collective, mossmann2006semiclassical, liu2007josephson, trimborn2009beyond,albiez2005direct, raghavan1999coherent, qiu2014measure}.

Here, by means of the \gls{TWA}, we study the dynamics after a \gls{MI} to \gls{SF} quench on a fully-connected lattice of generic number of sites $V$ and for a large number of bosons per lattice site. The choice of a fully-connected model is motivated by the analytical tractability allowed by its symmetries, by the fact that it represents an approximate description of a finite-dimensional system \cite{sciolla2010quantum, sciolla2011dynamical} and by potential experimental realizations with superconducting qubits \cite{nigg2017robust}. Under these assumptions, our work aim to capture the leading effects, neglecting possible corrections due to the finiteness of the lattice coordination number and of the density $\rho_0$, typically characterizing realistic systems. Remarkably, we reveal at short times the existence of a \gls{DPT} and at long times the relaxation of the system (for large $V$) to two qualitatively very different regimes on the two sides of the \gls{DPT}: for strong interactions the lattice sites are populated in an inhomogeneous way, a phenomenon that we refer to as \gls{MQST}, whereas for weak interactions the variables $\{\psi_j\}$ tend to split in two groups with phase difference $\pi$, an effect that we refer to as $\pi$-synchronization. Interestingly, we show that these features are intimately connected to the \glspl{FP} of the \gls{GPE}, finding that if and only if $V \ge 4$ there exists a peculiar manifold of infinitely many \glspl{FP} that enables the aforementioned synchronization. Our results are consistent with the idea that in high-dimensional lattices the system builds up long-range correlations at low effective temperatures (that is weak quenches). Furthermore, we show that no fine-tuning of the model parameters is needed for the occurrence of such $\pi$-synchronization. Indeed, we find that the long time asymptotic value of a suitable $\pi$-synchronization parameter $S$ decreases smoothly with the strength of a site-dependent disorder in what we refer to as a synchronization crossover.

The paper is organized as follows. In Sec.~\ref{sec: model} we write the system Hamiltonian, derive the corresponding \gls{GPE} and introduce the \gls{TWA}, describing the system evolution after a quench. In Sec.~\ref{section: The effects of conserved quantities} we present a powerful argument on the effects of conserved quantities (namely energy and total number of particles) on the non-equilibrium dynamics of the system, and find the mathematical condition underlining \gls{MQST}. In Sec.~\ref{sec: FPs} we find the \glspl{FP} of the \gls{GPE}, discovering the existence of the manifold of infinitely many \glspl{FP} for $V\ge4$. In Sec.~\ref{section: short time, linearization and DPT} we study the short time dynamics by means of the linear stability analysis of the \glspl{FP}, showing the \gls{DPT} and, on one of its sides, the appearance of the \gls{MQST}. In Sec.~\ref{section: long time and pi synchronization} we numerically implement the \gls{TWA} and investigate the long time dynamics, discovering the emergence under particular circumstances either of \gls{MQST} or of $\pi$-synchronization. From an analogy with liquid crystals \cite{prost1995physics}, we introduce the parameter $S$ to quantify the $\pi$-synchronization and in Sec.~\ref{section: competition between hopping strength and disorder} we study its robustness against the introduction of site-dependent disorder, finding the synchronization crossover. In Sec.~\ref{section: conclusions} we summarize our results and outline possible directions of further research. Finally, the appendices are mainly devoted to technical aspects and derivations. 
 
\section{Model}
\label{sec: model}
To describe the model, we start by deriving the dynamical equations of a system of interacting bosons on a fully-connected lattice, that is a lattice where particles can hop from any site to any other site with same tunneling strength. By means of a mean-field approximation, we are able to recast the Heisenberg equation of motion into a nonlinear and discrete \gls{GPE} of motion for classical coupled oscillators of variable length and phase. The \gls{TWA} is finally obtained considering an ensemble of initial stochastic classical fields.

\subsection{Hamiltonian}
The Bose-Hubbard model on a fully-connected lattice is characterized by the following Hamiltonian
\begin{equation}
H_{BH} = - \frac{J}{V} \mathop{\sum_{i,j = 1}}_{i \neq j}^{V} a_i^\dagger a_j + \frac{u}{2} \sum_{j=1}^{V} n_j (n_j - 1) - \mu \sum_{j=1}^{V} n_j \ ,
\label{eq: BH Hamiltonian}
\end{equation}
where $V$ is the number of lattice sites, $a_j^\dagger$ and $a_j$ are the  bosonic creation and annihilation operators at site $j$, respectively, satisfying the bosonic commutation relation $[a_i, a_j^\dagger] = \delta_{i,j}$, $n_j = a_j^\dagger a_j$ is the number operator associated to the $j$-th site, $J$ is the hopping strength for tunneling between any two sites (rescaled of a factor $V$ to guarantee extensivity), $u$ is the energy scale of the on-site two-body repulsive interaction ($u>0$) and $\mu$ the chemical potential setting the average number of particles in the system. In the following, the indices $i,j,k$ are assumed to run over all the sites $1,2,\dots,V$, unless differently specified. We denote by $N$  the total number of particles and by $\rho_0 = N/V$ the average number of particles per lattice site.

Relevant for the determination of both the equilibrium and the non-equilibrium properties of the system is the following dimensionless parameter
\begin{equation}
\eta = \frac{J}{u\rho_0} \ .
\label{eq: eta}
\end{equation}
It is well-known that, varying $\eta$ across a critical value $\eta_c^{eq}(\rho_0, \mu)$, the system undergoes an equilibrium phase transition between a \gls{SF} and a \gls{MI}, the former being characterized by long-range coherence and the latter by integer boson densities, existence of a gap for particle-hole excitation and zero compressibility \cite{fisher1989boson, freericks1994phase}. Within mean-field, at integer fillings and for $\rho_0 \gg 1$ the transition occurs at \cite{fisher1989boson, polkovnikov2002nonequilibrium}
\begin{equation}
\eta_c^{eq} \approx \frac{1}{4\rho_0^2} \ ,
\end{equation}
so that only a small interval $0 < \eta < \frac{1}{4\rho_0^2} \ll 1$ will correspond to a \gls{MI} ground state. As a consequence, switching $\eta$ from $\eta_i \approx 0$ to $\eta_f \sim 1$ at $t = 0$ corresponds in this limit to a \gls{MI} to \gls{SF} quench.
It should be emphasized that, in order to consistently work within the \gls{TWA} and capture the leading effects, we shall henceforth assume a large but finite $\rho_0 \gg 1$. Corrections to our model arise on the one hand from the deviation of realistic systems from the Bose-Hubbard model for large $\rho_0$ and on the other from the finite $\rho_0$ quantum effects beyond \gls{TWA} \cite{polkovnikov2003quantum}.

\subsection{Gross-Pitaevskii dynamical equations}

The dynamical equation for the bosonic annihilation operator at site $k$ is readily obtained within the Heisenberg formalism as ($\hbar=1$)
\begin{equation}
\frac{d a_k}{dt} = i [H_{BH}, a_k] \ .
\label{eq: Heisenberg dynamics}
\end{equation}
that reads (details in App.~\ref{App: Heiseberg equation})
\begin{equation}
\frac{d a_k}{d(it)} = + \frac{J}{V}\sum_{j=1}^{V} a_j - u n_k a_k \ .
\label{eq: dyn eq 3}
\end{equation}

At the mean-field level, for a large number of bosons per lattice site ($\rho_0 \gg 1$) and in the \gls{SF} regime ($\eta > \eta_c^{eq}$ with $\eta_c^{eq} \ll 1$ as explained above), a well-established approximation to approach Eq. \eqref{eq: dyn eq 3} reads \cite{polkovnikov2002nonequilibrium, smerzi1997quantum}
\begin{equation}
\begin{aligned}
& \langle n_k a_k \rangle \approx | \langle a_k \rangle |^2 \langle a_k \rangle \ , \\
& \langle n_k \rangle \approx \langle a_k^\dagger \rangle \langle a_k\rangle = |\langle a_k^\dagger \rangle |^2 \ ,
\end{aligned}
\label{eq: GP approximation}
\end{equation}
where $\langle\bullet\rangle$ denotes the expectation value. Because of the coupling with the environment prior to the quench, in general the system is not in an eigenstate of the total number of particles operator $\sum_{j=1}^{V} n_j$, and the expectation value $\langle a_k^\dagger \rangle$ is non-vanishing. We denote
\begin{equation}
\langle a_k \rangle = \psi_k = \sqrt{\rho_k}e^{i \theta_k} \ ,
\label{eq: rho-theta definition}
\end{equation}
where $\rho_k$ and $\theta_k$ are the squared modulus and the phase of $\psi_k$, respectively. From equation \eqref{eq: dyn eq 3} and under the approximation \eqref{eq: GP approximation} we obtain
\begin{equation}
\frac{d \psi_k}{d(it)} = + \frac{J}{V}\sum_{j=1}^{V} \psi_j - u |\psi_k|^2 \psi_k \ ,
\label{eq: dyn eq GP}
\end{equation}
that is a discrete and nonlinear \gls{GPE}. As detailed in App.~\ref{App: dynamical equations for rho_k and theta_k}, from Eq.~\eqref{eq: dyn eq GP} we can derive the dynamical equations for $\rho_k$ and $\theta_k$, reading
\begin{equation}
\begin{cases}
 \displaystyle\frac{d \sqrt{\rho_k}}{dt} &= \displaystyle \frac{J}{V} \sum_{j=1}^{V} \sqrt{\rho_j} \sin \left(\theta_k - \theta_j \right) \\
 \displaystyle \frac{d \theta_k}{dt} &= \displaystyle \frac{J}{V} \sum_{j=1}^{V} \sqrt{\frac{\rho_j}{\rho_k}} \cos \left(\theta_k -\theta_j\right) - u \rho_k \ .
\end{cases}
\label{eq: dyn eq GP - rho theta}
\end{equation}

We define the following complex dynamical order parameter
\begin{equation}
\Psi = re^{i\phi} = \frac{1}{V} \sum_{j = 1}^{V} \sqrt{\rho_j} e^{i\theta_j} \ ,
\label{eq: order parameter}
\end{equation}
whose modulus and phase are denoted $r$ and $\phi$, respectively. Similarly to what is tipically done for the Kuramoto model for classical coupled oscillators \cite{acebron2005kuramoto}, considering the real and the imaginary part of $re^{i \left(\phi - \theta_k\right)} = \frac{1}{V} \sum_{j = 1}^{V} \sqrt{\rho_j} e^{i \left(\theta_j - \theta_k\right)}$, we readily find
\begin{equation}
\begin{aligned}
& r\cos(\phi - \theta_k) = \frac{1}{V} \sum_{j = 1}^{V} \sqrt{\rho_j} \cos(\theta_j - \theta_k) \ ,\\
& r\sin(\phi - \theta_k) = \frac{1}{V} \sum_{j = 1}^{V} \sqrt{\rho_j} \sin(\theta_j - \theta_k)  \ ,\\
\end{aligned}
\end{equation}
so that Eq.~\eqref{eq: dyn eq GP - rho theta} can be compactly rewritten as
\begin{equation}
\begin{cases}
\displaystyle \frac{d \sqrt{\rho_k}}{dt} &= \displaystyle J r \sin \left(\theta_k - \phi \right) \\[10pt]
\displaystyle \frac{d \theta_k}{dt} &=\displaystyle  \frac{J r}{\sqrt{\rho_k}}\cos \left(\theta_k -\phi\right) - u \rho_k \ ,
\end{cases}
\label{eq: dyn eq GP - rho theta - compact - no disorder.0}
\end{equation}
where we stress that $r$ and $\phi$ are in general time-dependent, evolving consistently with all the variables $\{\rho_j,\theta_j\}$, accordingly to Eq.~\eqref{eq: order parameter}. Being the system isolated, the average number of particles per lattice sites $\rho_0 = \frac{1}{V} \sum_{j = 1}^{V} \rho_j$ is a conserved quantity of Eq.~\eqref{eq: dyn eq GP - rho theta - compact - no disorder.0}.

Expressing the time $t$ in units of $\frac{\hbar}{u \rho_0}$ and $\rho_k$ in units of $\rho_0$, Eq.~\eqref{eq: dyn eq GP - rho theta - compact - no disorder.0} is rewritten as
\begin{equation}
\begin{cases}
 \displaystyle\frac{d \sqrt{\rho_k}}{dt} = \eta r \sin \left(\theta_k - \phi \right) \\[10pt]
 \displaystyle\frac{d \theta_k}{dt} = \eta \frac{r}{\sqrt{\rho_k}}\cos \left(\theta_k -\phi\right) - \rho_k \ ,
\end{cases}
\label{eq: dyn eq GP - rho theta - compact - no disorder}
\end{equation}
where $\eta$ is the dimensionless hopping strength defined in Eq.~\eqref{eq: eta}. Importantly, expressing $\rho_k$ in units of $\rho_0$, the average of $\rho_k$ over the sites is renormalized to 1, that is $\rho_0 = \frac{1}{V}\sum_{j=1}^{V} \rho_j = 1$. Similarly, $r$ will assume values in $(0,1)$. The \gls{GPE} \eqref{eq: dyn eq GP - rho theta - compact - no disorder} consists of a system of ordinary differential equations for the $2V$ real variables $\{\sqrt{\rho_k}, \theta_k\}$. We call \emph{configuration} the $2V$-dimensional set of variables $\{\sqrt{\rho_k}, \theta_k\}$ associated to the state of the system and \emph{phase space} the $2V$-dimensional space in which the configurations live. Finally, we observe that the approximation \eqref{eq: GP approximation} corresponds to considering a classical Hamiltonian
\begin{equation}
H_{CL} = V\left(- \eta r^2 + \frac{1}{2} \frac{1}{V} \sum_{j=1}^{V} \rho_j^2 \right) \ .
\label{eq: H_CL}
\end{equation}

\subsection{Quench, \gls{TWA} and system initialization}
\begin{figure}[t]
	\begin{center}
		\includegraphics[width=\linewidth]{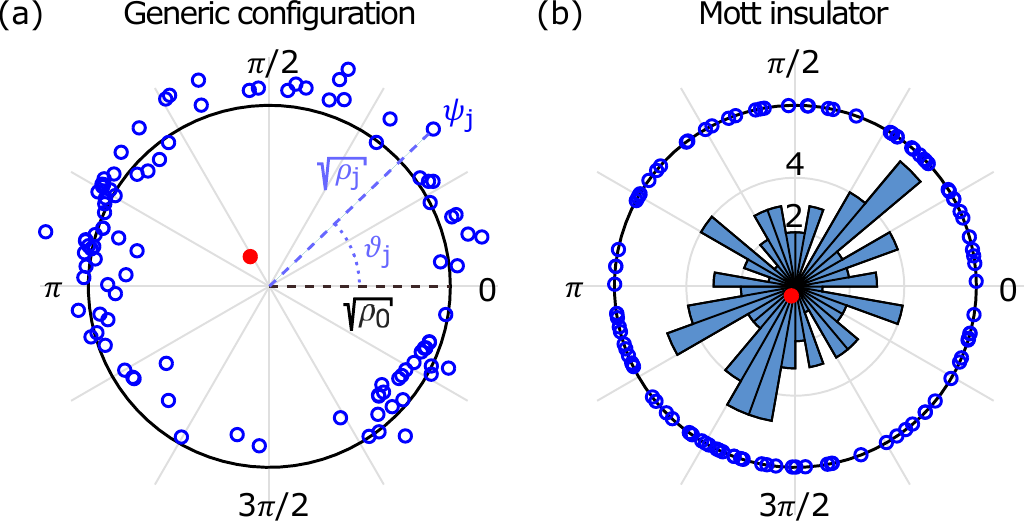}\\
	\end{center}
	\vskip -0.5cm \protect\caption[]
	{(color online) Graphical representation of the system configuration for $V = 100$ lattice sites. In the complex plane, the blue markers represent the mean-field bosonic variables $\{\psi_j=\sqrt{\rho_j}e^{i\theta_j}\}$ (one marker per each site), a red dot represents the complex dynamical order parameter $\Psi=r e^{i\phi}$ and a black circle of radius $\sqrt{\rho_0}=1$ is drawn as a reference. (a) Generic configuration, for which the $\rho_j$ are spread around $\rho_0$ and $r \neq 0$. (b) Example of initial configuration ($t=0$) for a \gls{MI} to \gls{SF} quench [Eq.~\eqref{eq: MI}]. A polar histogram with $10^\circ$ wide bins illustrates the distribution of the phases $\{\theta_j\}$. The proximity of the red dot to the origin of the plane reflects the fact that $r \ll 1$ for a \gls{MI} and $V\gg1$.}
	\label{fig:complex_representation}
\end{figure}

To study the system dynamics in the non-equilibrium regime, we adopt the prototypical quench procedure, consisting of a sudden change of the Hamiltonian at time $t = 0$ from $H_i$ to $H_f$ \cite{sengupta2004quench, moeckel2008interaction, chin2010feshbach, schiro2010time, sciolla2010quantum, sciolla2011dynamical, sciolla2013quantum}. Thanks to the high degree of isolation of the system achievable on the experimental timescales \cite{langen2015ultracold}, this procedure enables to investigate an almost-isolated system initialized in the ground state of the Hamiltonian $H_i$ and evolving under the Hamiltonian $H_f$ for $t>0$. We shell mostly focus on (but not limit ourselves to) the \gls{MI} to \gls{SF} quench, corresponding to a change of $\eta$ from $\eta_i \approx 0$ to $\eta_f \sim 1$ at $t = 0$ (we recall that $\eta_c^{eq} \ll 1$). In this case the \gls{TWA} considers initial fields characterized by uniform density and stochastic phases \cite{polkovnikov2002nonequilibrium}
\begin{equation}
\begin{cases}
\rho_j(t=0) = \rho_0 \\
\theta_j(t=0) = U_j  \ ,
\end{cases}
\label{eq: MI}
\end{equation}
$\{U_j\}$ being independent uniform random numbers between $0$ and $2\pi$. Importantly, for a large number of sites $V$ the \gls{MI} of \eqref{eq: MI} is characterized by $r \sim 1/\sqrt{V} \xrightarrow{V \rightarrow \infty} 0$. A graphical representation of the system mean-field state is shown in Fig.~\ref{fig:complex_representation}, where $V$ blue markers in the complex plane represent the variables $\{\psi_j=\sqrt{\rho_j}e^{i\theta_j}\}$ (one marker per each site), the red marker represents $\Psi=re^{i\phi}$ and a black circle of radius $\sqrt{\rho_0}$ is shown as a reference. Additionally, a polar histogram of the phases $\{\theta_j\}$ with bin width $10^\circ$ is possibly displayed (b). To better interpretate such a representation, it is worth to stress that $r$ represents the distance of the red marker from the origin of the complex plane. The fact that for large $V$ the \gls{MI} \eqref{eq: MI} is characterized by $r\approx 0$ reflects in its representation (b) into the red marker being close to the origin.

Having mapped the deterministic quantum evolution onto a stochastic classical evolution, it is possible to approximate the quantum expectation value of a normally ordered physical observable $f(\{a_j, a_j^\dagger\})$ at any $t>0$ as \cite{polkovnikov2002nonequilibrium}
\begin{equation}
\langle f(\{a_j, a_j^\dagger\}) \rangle \approx
\langle f(\{\psi_j, \psi_j^*\}) \rangle_{\text{random } \{\theta_j(t = 0)\}} \ ,
\label{eq: exp value}
\end{equation}
where the RHS denotes average over the evolutions at time $t>0$ corresponding to different realizations of the initial random phases of \eqref{eq: MI}. This procedure goes under the name of \gls{TWA} and is exact up to an error of order $1/\rho_0$ \cite{polkovnikov2003quantum}. For simplicity and without ambiguity, in the following we shall however adopt only the notation $\langle \dots \rangle$, that has to be interpreted in the sense of Eq.~\eqref{eq: exp value}. To evaluate the RHS of Eq.~\eqref{eq: exp value}, we aim to study analytically the dynamics generated by the \gls{GPE} for a generic initial configuration of the phases $\{\theta_j\}$. Such study is carried on within the framework of Dynamical System Theory in Secs.~\ref{section: The effects of conserved quantities}, \ref{sec: FPs} and \ref{section: short time, linearization and DPT}. From a computational point of view instead we perform an average over the numerical solutions of the \gls{GPE} obtained for a large number of simulations, each one for different random initial phases, as done in Secs.~\ref{section: long time and pi synchronization} and \ref{section: competition between hopping strength and disorder}.

\section{Effects of conserved quantities on the non-equilibrium dynamics}
\label{section: The effects of conserved quantities}
On the experimentally relevant timescales our system can be considered isolated \cite{hofferberth2007non, cramer2008exploring, flesch2008probing}, meaning that the non-equilibrium dynamics of the system will be constrained by the the presence of conserved quantities. Conservation of energy is for instance preventing the motion of the system from the ground state of the Hamiltonian $H_i$ (preceding the quench) to the ground state of the Hamiltonian $H_f$ (following the quench). In this preliminary Section, we show that relevant information on the non-equilibrium dynamics of the system can be easily obtained from the conservation of the total number of particles and of the energy \eqref{eq: H_CL}, that reads
\begin{equation}
- \eta r^2 + \frac{1}{2V} \sum_{j=1}^{V} \rho_j^2 = E \ , \label{eq: conserved quantities 1}
\end{equation}
where $E$ is a constant depending on the initial condition. Writing $\rho_j$ as $\rho_j = \rho_0 + \delta_j$, the conservation of the total number of particles reads $\sum_{j=1}^{V} \delta_j = 0$. Furthermore, in the particularly interesting case of initial homogeneous density ($\delta_j = 0$), denoting $r_0 = r(t = 0)$, we get at initial time $t = 0$ that $E = \frac{1}{2} - \tau r_0^2$, so that Eq.~\eqref{eq: conserved quantities 1} reduces to
\begin{equation}
\frac{1}{V}\sum_{j=1}^{V} \left(\rho_j - \rho_0\right)^2 = 2 \eta (r^2 - r_0^2) \ , \label{eq: consistency conditions bis 1}
\end{equation}
conveying important information on the system non-equilibrium dynamics. First, since the LHS of Eq.~\eqref{eq: consistency conditions bis 1} is positive definite, for all times $t>0$ we have
\begin{equation}
r(t) \ge r_0  \ .
\label{eq: condition on r for conservation}
\end{equation}
An immediate consequence is that, being $r = 1$ the largest possible $r$, a system initialized close to the superfluid phase (that is with $r \approx 1$) will remain close to the superfluid phase (namely with $r \approx 1$). The second implication of Eq.~\eqref{eq: consistency conditions bis 1} is that an increase of $r$ must be accompanied by a spread of the $\{\rho_j\}$ around their mean value $\rho_0 = 1$ [as happens in Fig.~\ref{fig:complex_representation}(a)]. The goal of the next Section is to unveil the conditions under which such growth of $r$ occurs. From Eq.~\eqref{eq: consistency conditions bis 1} we finally notice that, in the recurrent case of $r_0 \approx 0$, $r$ is a direct measure of the inhomogeneity of population among the sites, generalizing to an arbitrary $V$ what in the $V = 2$ case is called population imbalance. Therefore, a finite $\langle r \rangle$ corresponds to \gls{MQST}.

\section{Fixed points}
\label{sec: FPs}
Aiming to study the dynamics of the system when initialized with homogeneous density and generic phases $\{\theta_j\}$, we start looking for the \glspl{FP} of the \gls{GPE}, that are the configurations that are preserved in time. In fact, in the proximity of a \gls{FP} the short time dynamics can be studied by means of a linearization of the \gls{GPE}. Furthermore, the behavior of the system is intimately related to the \glspl{FP} even at long times: a manifold of infinitely many \glspl{FP} in the phase space, that is peculiar of $V \ge 4$, allows the asymptotic $\pi$-synchronization of the phases $\{\theta_j\}$, as we will show in Sec.~\ref{section: long time and pi synchronization}. Being of crucial importance for the determination of the dynamical properties of the system, in this Section we systematically find and chategorize the \glspl{FP}, assigning names and acronymes to the most relevant of them, that will be extensively adopted in the remainder. A clear intuition of the \glspl{FP} is given by their systematic representation in Fig.~\ref{fig:SCs and phase space}, to which the reader may refer throughout this Section.

To find the \glspl{FP} we conveniently adopt a definition for stationarity that allows a common rotation of all the phases $\{\theta_j\}$ at some constant rate $\Omega$ (that can in fact always be removed with a proper gauge transformation, as shown in App.~\ref{App: Heiseberg equation}). Therefore, a configuration that fulfills for some $\Omega$ the following stationarity conditions for all sites $k$
\begin{align}
&\displaystyle\frac{d \sqrt{\rho_k}}{dt} = \displaystyle \eta r \sin \left(\theta_k - \phi \right) = 0 
\label{eq: stationarity condition 1} \ ,\\
&\displaystyle \frac{d \theta_k}{dt} =\displaystyle \frac{\eta r}{\sqrt{\rho_k}}\cos \left(\theta_k -\phi\right) - \rho_k = \Omega \ ,
\label{eq: stationarity condition 2}
\end{align}
shall be called a \gls{FP} of the \gls{GPE} \eqref{eq: dyn eq GP - rho theta - compact - no disorder}. Clearly, being $\Omega$ site-independent, it follows from Eq.~\eqref{eq: stationarity condition 2} that $\phi(t) = \phi(0) + \Omega t$. With a suitable choice of the reference frame we set $\phi(0) = 0$, so that at $t=0$ Eq.~\eqref{eq: order parameter} reads
\begin{equation}
\frac{1}{V} \sum_{j = 1}^{V} \sqrt{\rho_j} e^{i\theta_j} = r \ , \label{eq: consistency conditions}
\end{equation}
From Eq.~\eqref{eq: stationarity condition 1} we find that only two kinds of \glspl{FP} are possible: the ones with $\sin(\theta_k) = 0$ and the ones with $r = 0$. We address these 2 classes of \glspl{FP} separately. In the following, all the relevant \glspl{FP} are defined up to a site permutation (as natural for a fully-connected model) and a rotation of the reference frame.

\subsection{Superfluid and $\pi$-aligned fixed points}
\label{section: superfluid and pi-aligned SCs}
The first class of \glspl{FP} is characterized by $\theta_k \in \{0, \pi\}$. The simpler \gls{FP} with such property is the \gls{SFC}, with homogeneous density $\rho_k = \rho_0$ and equal phases, that is
\begin{equation}
\begin{cases}
\rho_k = 1\\
\theta_k = 0 \ ,
\end{cases}
\label{eq: stationarity condition - aligned}
\end{equation}
for which we get $r = 1$ and $\Omega = \eta - 1$. A graphical representation of the \gls{SFC} is shown in Fig.~\ref{fig:SCs and phase space}(f). Maximizing $r$, the \gls{SFC} is the ground state of the semiclassical Hamiltonian \eqref{eq: H_CL}. In the non-equilibrium regime, the system will in general be far from the \gls{SFC}.

For an even $V$, a second relevant \gls{FP} in this class is the one with homogeneous density, one half of the phases equal to $0$ and the other equal to $\pi$, reading
\begin{equation}
\begin{cases}
\rho_k = 1 \\
\theta_k = k\pi  \ ,
\end{cases}
\label{eq: stationarity condition - SPAC}
\end{equation}
for which $r = 0$ and $\Omega = -1$, that we call \gls{SPAC} and whose graphical representation is shown in Fig.~\ref{fig:SCs and phase space}(d). Experimentally, it is possible to initialize a cold atoms system to the \gls{SPAC} applying short pulses to the condensate \cite{polkovnikov2002nonequilibrium}. In such case the short time dynamics can thus be studied linearizing the \gls{GPE} around the \gls{SPAC}.

Other possible \glspl{FP} in this class have a fraction $\alpha \neq 1/2$ of sites with phase $0$ and the remaining fraction $1-\alpha$ with phase $\pi$ and will generically be referred to as $\pi$-aligned configurations (one example is shown in Fig.~\ref{fig:SCs and phase space}(e)). The relevance for our study of these configurations is limited.

\subsection{$r = 0$ fixed points}
The second class of \glspl{FP} is characterized by $r = 0$, for which the condition \eqref{eq: stationarity condition 2} reads $\rho_k = -\Omega$. Consequently, Eq.~\eqref{eq: consistency conditions} reads
\begin{equation}
\sum_{j = 1}^{V} e^{i\theta_j} = 0 \ .
\label{eq: stationarity condition - r = 0}
\end{equation}
Eq.~\eqref{eq: stationarity condition - r = 0} has in general many solutions (namely infinite if and only if $V \ge 4$). Of course, the aforementioned \gls{SPAC} is one of them, in fact being the only \gls{FP} satisfying at the same time $r = 0$ and $\theta_k - \theta_j \in \{0, \pi\}$.

\begin{figure}[h]
	\begin{center}
		\includegraphics[width=\linewidth]{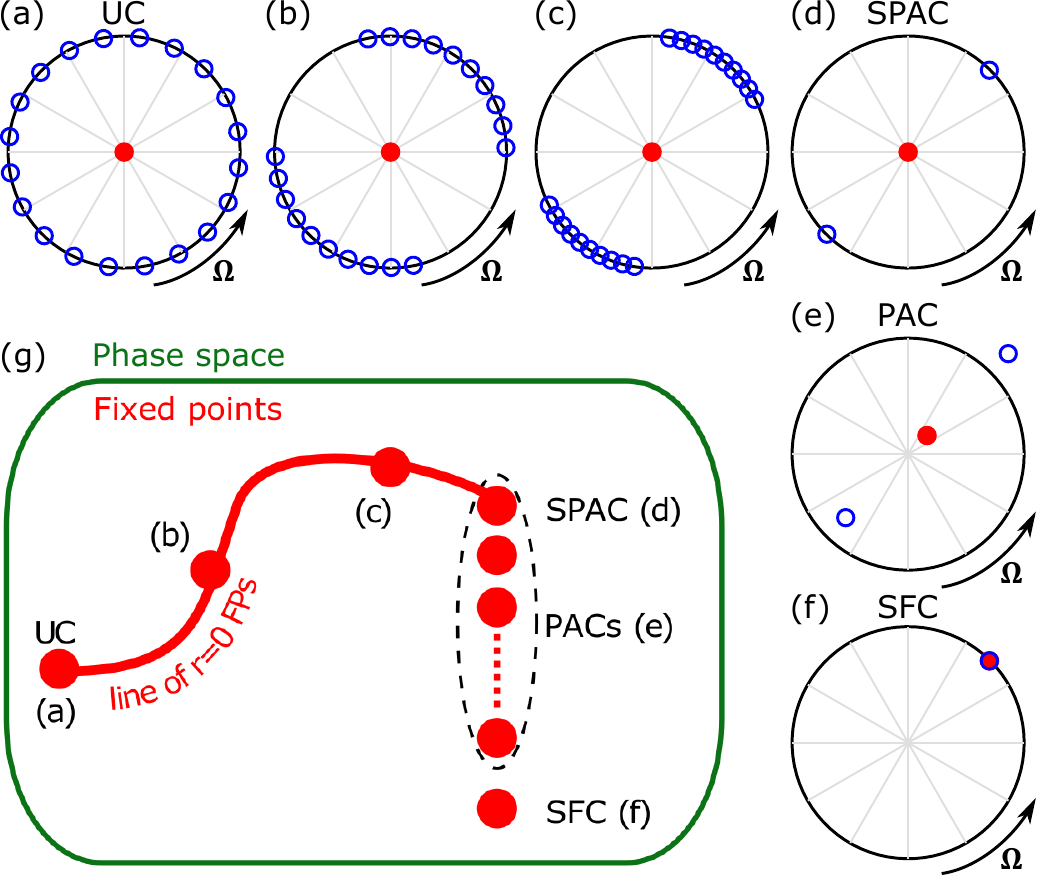}\\
	\end{center}
	\vskip -0.5cm \protect\caption[]
	{(color online) Schematic representation of the representative \glspl{FP} of the \gls{GPE} \eqref{eq: dyn eq GP - rho theta - compact - no disorder} for an even $V\ge4$. For $V = 20$ sites we show the \gls{UC} (a), two other \glspl{FP} with $r=0$ (b,c), the \gls{SPAC} (d), one $\pi$-aligned configuration (e) and the \gls{SFC} (f). The circular arrows indicate that the phases of a \gls{FP} are in general rotating at some constant rate $\Omega$. (g) The $r = 0$ \glspl{FP} constitute a manifold in the phase space that ranges from the \gls{UC} to the \gls{SPAC}. Notice that also for an odd $V\ge5$ there is an analogue manifold of infinitely many $r = 0$ \glspl{FP}, just lacking of the \gls{SPAC}.}
	\label{fig:SCs and phase space}
\end{figure}

Certainly the most important \gls{FP} for our study is the configuration defined for $V \ge 3$ by
\begin{equation}
\begin{cases}
\rho_k = 1\\
\theta_k = \frac{2 \pi}{V}k  \ ,
\end{cases}
\label{eq: stationarity condition - UC}
\end{equation}
that we call \gls{UC}, where the word \emph{uniform} is used to stress the uniform spacing $2\pi/V$ of the phases. A graphical representation of the \gls{UC} is shown in Fig.~\ref{fig:SCs and phase space}(a). Importantly, we observe that, in the infinite dimensional limit ($V \rightarrow \infty$) and for a proper permutation of the sites, the \gls{UC} coincides with the \gls{MI} \eqref{eq: MI}, since a number $V\rightarrow\infty$ of uniform random phases in $(0,2\pi)$ is equivalent to $V$ equispaced phases over the same interval. For a large but finite $V \gg 1$, random noise will instead make a generic \gls{MI} configuration \eqref{eq: MI} different from the \gls{UC} \eqref{eq: stationarity condition - UC} but close to it. This observation crucially reflects into the fact that for a \gls{MI} to \gls{SF} quench and $V \gg 1$, the system is initialized in the proximity of the \gls{UC} and the dynamics at short times can thus be studied linearizing the \gls{GPE} around the \gls{UC}.

Finally, we notice that for $V \ge 4$ the condition \eqref{eq: stationarity condition - r = 0} defines an infinity of \glspl{FP} (e.g.~the ones shown in Fig.~\ref{fig:SCs and phase space}(b,c) for $V = 20$), constituting a $(V-3)$-dimensional manifold in the phase space and of which the \gls{UC} and (if $V$ is even) the \gls{SPAC} are part, as schematically shown in Fig.~\ref{fig:SCs and phase space}(g). For $V = 4$ such manifold is a line, and can be represented parametrically by the following \gls{FP}
\begin{equation}
\begin{cases}
\rho_k = 1 \\
\theta_1 = +\frac{\pi}{2} - \frac{\Delta}{2}\\
\theta_2 = -\theta_1 \\
\theta_3 = \theta_1 + \pi \\
\theta_4 = \theta_2 + \pi  \ ,
\end{cases}
\label{eq: DC}
\end{equation}
that we call \gls{DC} as it depends on the parameter $\Delta$ and that ranges continuously from the \gls{UC} ($\Delta=\pi/2$) to the \gls{SPAC} ($\Delta=\pi$). The importance of the \gls{DC} lies in the fact that it enables us to carry on analytical calculations along the manifold of the $r = 0$ \glspl{FP} for $V = 4$, with generalizations to $V>4$, for which we instead focus on the \gls{UC} and the \gls{SPAC} only.

\begin{figure}[t]
	\begin{center}
		\includegraphics[width=\linewidth]{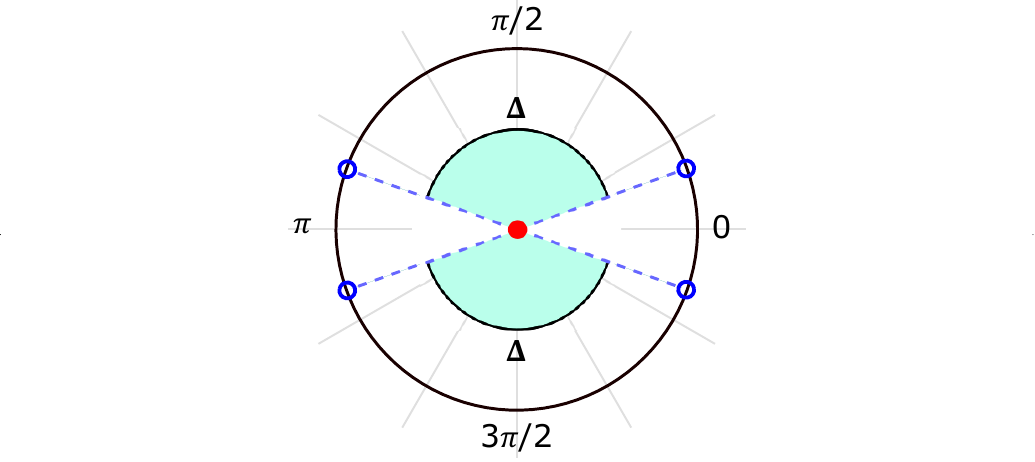}\\
	\end{center}
	\vskip -0.5cm \protect\caption[]
	{(color online) Graphical representation of the \gls{DC}, defined for $V = 4$ lattice sites in \eqref{eq: DC}. The \gls{DC} is a parametric \gls{FP} that, depending on the value of the parameter $\Delta \in (\pi/2, \pi)$, spans the entire manifold of $r = 0$ \glspl{FP}, ranging from the \gls{UC} (for $\Delta = \pi/2$) to the \gls{SPAC} (for $\Delta = \pi$).}
	\label{fig: DC}
\end{figure}

\section{Short time behavior: the dynamical phase transition}
\label{section: short time, linearization and DPT}

Having argued that at time $t = 0$ the system can be initialized either in the proximity of the \gls{UC} or of the \gls{SPAC}, we can now proceed with the study of the dynamics of a system initialized in the proximity of the \glspl{FP} thanks to a linearization of the \gls{GPE}. Here, after briefly reviewing the instructive $V = 2$ case, we extend it to all possible $V \ge 4$, thus covering also to the large dimensional limit $V \gg 1$. In the framework of The Dynamical Systems Theory, by means of a diagonalization of the Jacobian matrix $J$ (not to be confused with the dimensional hopping strength) associated to the linearized \gls{GPE} in the neighborhood of the most relevant \glspl{FP}, we explore the non-equilibrium dynamics at short times after the quench. Particularly, this is relevant for a \gls{SF} to \gls{MI} quench for $V \gg 1$, for which the system is initialized in the proximity of the \gls{UC} \eqref{eq: MI} and for an initialization of the system to the \gls{SPAC} (achievable in cold atoms applying short pulses to the condensate \cite{polkovnikov2002nonequilibrium}). Looking at the eigenvalues of $J$, we find two regions of the parameter space corresponding to two qualitatively very different behaviors of the system in what can be called a dynamical phase transition.

We start by linearizing the \gls{GPE} \eqref{eq: dyn eq GP - rho theta - compact - no disorder}. To this purpose we introduce the $2V$-dimensional column vector
\begin{equation}
\vec{y} = \left(\theta_1, \theta_2, \dots, \theta_V, \sqrt{\rho_1}, \sqrt{\rho_2}, \dots, \sqrt{\rho_V} \right)^T \ ,
\label{eq: state vector}
\end{equation}
that describes the state of the system at the mean-field level. The Jacobian $J$ associated to the \gls{GPE} \eqref{eq: dyn eq GP - rho theta - compact - no disorder} is the $2V\times2V$-dimensional matrix with entries
\begin{equation}
J_{n, m} = \frac{\partial}{\partial y_m} \left(\frac{dy_n}{dt}\right) \quad n, m = 1, 2, \dots, 2V \ .
\label{eq: Jacobian.0}
\end{equation}

If the system is initialized in a state $\vec{y}(0)$ in the proximity of a \gls{FP} $\vec{y}^{FP}$, the solution of the linearized \gls{GPE} reads \cite{strogatz2018nonlinear}
\begin{equation}
\vec{y}(t) = \vec{y}^{FP} + e^{Jt} (\vec{y}(0) - \vec{y}^{FP}) \ ,
\label{eq: linearized solution}
\end{equation}
where the Jacobian matrix $J$ is evaluated in $\vec{y}^{FP}$. From Eq.~\eqref{eq: linearized solution} it follows that  the dynamics of a system is determined by the eigenvalues $\{\lambda_n\}$ of $J$ \cite{strogatz2018nonlinear}. For \glspl{FP} with $r = 0$ (such as the \gls{UC}, the \gls{SPAC} and the \gls{DC}) the latter turns out to read (see details in App.~\ref{app: the linearization procedure})
\begin{equation}
\begin{aligned}
J_{j, k} &=  - \frac{\eta}{V} \sin \left(\theta_k - \theta_j\right) \ ,\\
J_{j+V, k+V} &= -\frac{\eta}{V} \sin(\theta_k - \theta_j) \ ,\\
J_{j+V, k} &= - \frac{\eta}{V} \cos \left(\theta_k - \theta_j\right) \ ,\\
J_{j, k+V} &= + \frac{\eta}{V} \cos(\theta_k - \theta_j) - 2\delta_{k,j} \ ,
\end{aligned}
\label{eq: Jacobian, r = 0}
\end{equation}
from which we readily find that $\Tr{J} = 0$, meaning that the real parts of the Jacobian eigenvalues cannot be all positive or all negative, as expected for a conservative system. Rather, depending on the considered \gls{FP} and on $\eta$, only the following two situations are possible
\begin{itemize}
	\item All the eigenvalues $\{\lambda_n\}$ are purely imaginary (possibly 0), that is the \gls{FP} is a \emph{linear center} of the dynamics: the solution of the linearized equations is a state cycling periodically and close by the \gls{FP} when initialized in its proximity;
	
	\item Some eigenvalues have positive real part and some others have negative real part, that is the \gls{FP} is a \emph{saddle} of the dynamics: the solution of the linearized equations is a state moving exponentially fast apart from the \gls{FP} when initialized in its surroundings (because of random noise on the initial condition we exclude the possibility of system initialization exactly along a linear combination of eigenvectors associated to the eigenvalues with negative real part only).
\end{itemize}
Importantly, the \gls{FP} in the former case is termed a \emph{linear} center, since the above arguments on the eigenvalues are exact only for the linearized \gls{GPE}. A priori, a \emph{linear} center is not necessarily a \emph{nonlinear} center, that is the nonlinearities of the \gls{GPE} can make the system eventually move away from the \gls{FP} at long times even if the latter is a linear center \cite{strogatz2018nonlinear}. Interestingly, for a conservative system, a linear center is also a nonlinear center if it is \emph{isolated} (meaning that it is not part of a continuum of \glspl{FP}). Since if the \gls{FP} is a saddle (linear center) the system will (will not) drift away exponentially fast from it, with some abuse of nomenclature we will often refer to it as being \emph{stable} (\emph{unstable}).

Exploiting the presence of conserved quantities, in Sec.~\ref{section: The effects of conserved quantities} we showed that a system initialized in the proximity of the \gls{SFC} always orbits closely around it, meaning that the \gls{SFC} is a nonlinear center of the dynamics for any value of $\eta>0$.  With the above argument we can thus deduce that the \gls{SFC} is an isolated \gls{FP} (as indeed found in Sec.~\ref{sec: FPs}) and that the associated eigenvalues of $J$ are purely imaginary (as explicitly verified for completeness in App. \ref{app: the linearization procedure}). Instead, for a given \gls{FP} with $r = 0$, it turns out that there exists a critical value $\eta_c^{FP}$ of the dimensionless hopping strength such that the \gls{FP} is a saddle for $0<\eta<\eta_c^{FP}$ and a linear center for $\eta>\eta_c^{FP}$. This feature, known as bifurcation in Dynamical Systems Theory, leads to two qualitatively very different behaviors for a system initialized in the proximity of the considered \gls{FP} (e.g.~\gls{UC} for a \gls{MI} to \gls{SF} quench for $V \gg 1$) depending on $\eta \lessgtr \eta_c^{FP}$, that is a \gls{DPT} \cite{eckstein2009thermalization, sciolla2010quantum, sciolla2011dynamical, sciolla2013quantum}. In this Section, we exactly diagonalize the Jacobian matrix $J$ and find $\eta_c^{FP}$ for the \gls{DC} ($V=4$), the \gls{UC} (for $V \ge 3$) and the \gls{SPAC} (for an even $V$), thus locating the \gls{DPT}.

\subsection{A short review of the $V = 2$ case}
\label{subsec: V = 2 review}
\begin{figure}[t]
	\centering
	\includegraphics[width=\linewidth]{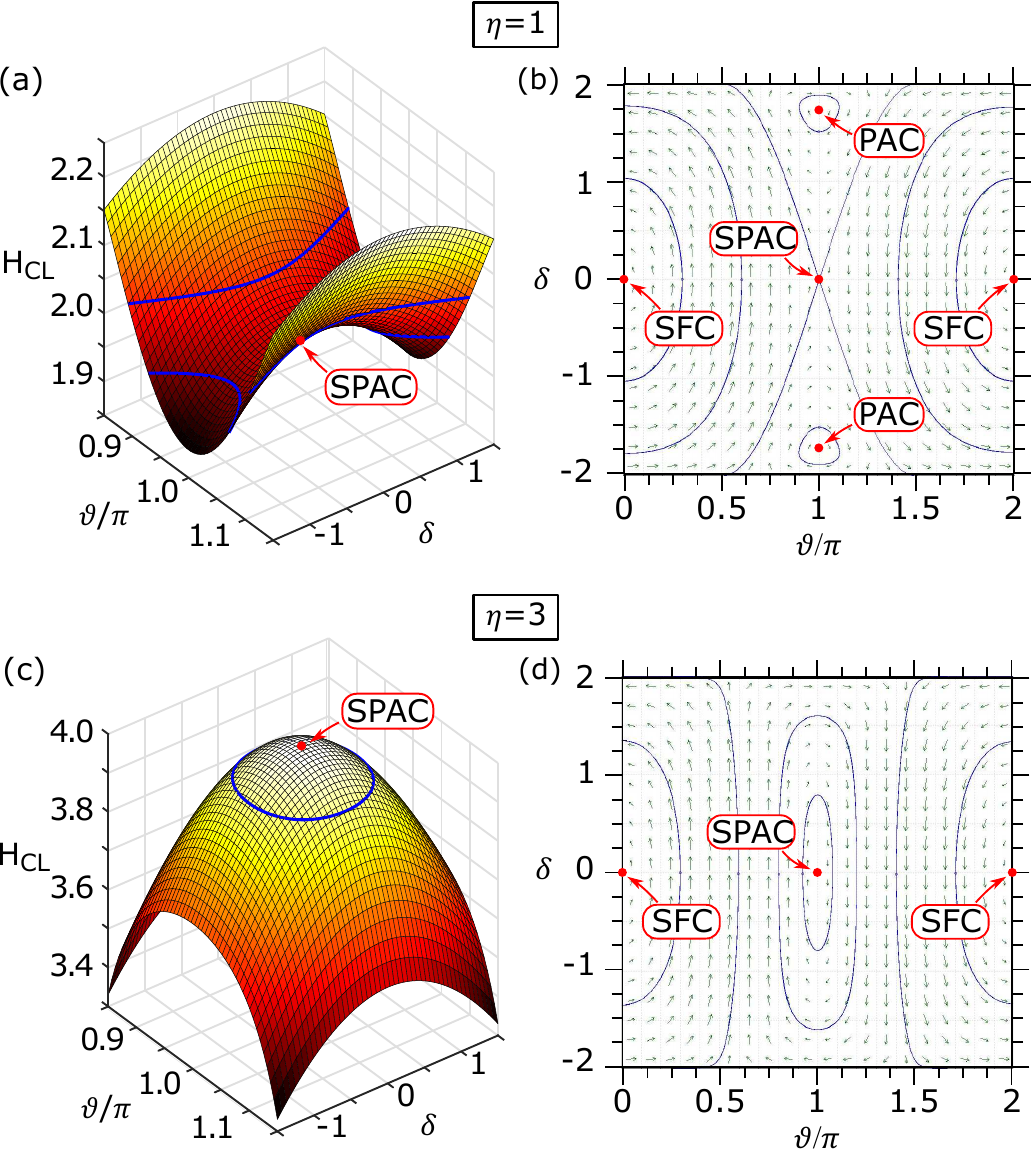}
	\caption{Dynamics of the two-site model in the phase space with coordinates $\theta = \theta_1 - \theta_2$ and $\delta = \rho_1 - \rho_2$. (a) semiclassical energy \eqref{eq: H_CL} in the sourroundings of the \gls{SPAC} ($\delta = 0, \ \theta = \pi$) together with some relevant trajectories (in blue) for $\eta = 1 < \eta_c^{SPAC} = 2$. (b) Phase portrait for $\eta = 1$: we show the \glspl{FP} (red dots), some relevant non-equilibrium trajectories (blue lines) and the flow (green arrows) associated to the \gls{GPE} \eqref{eq: dyn eq GP 2 sites - reduced}. (c,d) Energy landscape and phase portrait for $\eta = 3 > \eta_c^{SPAC}$. The \gls{SPAC} is either a saddle or a maximum of the semiclassical Hamiltonian depending on $\eta < \eta_c$ (a) or $\eta>\eta_c$ (c) and corresponding to a saddle (b) and a nonlinear center (d) of the dynamics, respectively. Instead, the \gls{SFC} ($\theta = \delta = 0$) and the $\pi$-aligned configurations are nonlinear centers for any $\eta>0$ and for $0<\eta<\eta_c$ respectively (b,d). Noticeably, the isolation (from the other \glspl{FP}) of the \gls{SPAC} is in stark contrast with the $V\ge4$ case.}
	\label{fig:2sites}
\end{figure}

Before addressing the higher $V$ case, it is useful to recall the results of the two-site system (that is a bosonic dimer) \cite{milburn1997quantum, smerzi1997quantum, franzosi2000quantum, gati2007bosonic, graefe2008mean, longhi2011optical, albiez2005direct, raghavan1999coherent, chuchem2010quantum}, with further details given in App.~\ref{app: 2sites}. Exploiting the constraint of conservation of the total number of particles ($\rho_1 + \rho_2 = 2\rho_0 = 2$), one can reduce the \gls{GPE} \eqref{eq: dyn eq GP - rho theta - compact - no disorder} to
\begin{equation}
\begin{cases}
\displaystyle\frac{\partial \theta}{\partial t} = - \delta - \frac{\eta}{2} \frac{\delta}{\sqrt{1 - \frac{\delta^2}{4}}} \cos \theta \\[10pt]
\displaystyle\frac{\partial \delta}{\partial t} = 2 \eta \sqrt{1 - \frac{\delta^2}{4}} \sin \theta \ ,
\end{cases}
\label{eq: dyn eq GP 2 sites - reduced}
\end{equation}
where $\delta = \rho_1 - \rho_2$ is the population imbalance and $\theta = \theta_1 - \theta_2$ is the phase difference between the two sites. The Jacobian eigenvalues associated to the \gls{GPE} \eqref{eq: dyn eq GP 2 sites - reduced} for the various \glspl{FP} are
\begin{equation}
\begin{aligned}
& \lambda_{1,2}^{SFC} = \pm 2 i \sqrt{ \frac{\eta}{2} \left(\frac{\eta}{2} + 1\right)} \ , \\
& \lambda_{1,2}^{SPAC} = \pm 2 \sqrt{\frac{\eta}{2} \left(1 - \frac{\eta}{2}\right)} \ , \\
& \lambda_{1,2}^{PAC_\pm} = \pm 2 i \sqrt{\frac{\eta}{2} \left(1 + \frac{4}{\eta^2} \right)} \ ,
\end{aligned}
\end{equation}
where PAC$_{+}$ and PAC$_{-}$ are two possible $\pi$-aligned configurations existing only for $\eta < \eta_c = 2$ and having $r \neq 0$. Being $\lambda_{1,2}^{SFC}$ and $\lambda_{1,2}^{PAC_\pm}$ purely imaginary for any $\eta>0$, the \gls{SFC} and (when existing) the PAC$_{\pm}$ are linear centers of the dynamics. Conversely, the \gls{SPAC} presents a double nature depending on the value of $\eta$: for $0 < \eta < \eta_c = 2$ it is a saddle of the dynamics ($\lambda_{1}^{SPAC} > 0$ and $\lambda_{2}^{SPAC} < 0$) whereas for $\eta > \eta_c = 2$ it is a linear center of the dynamics ($\lambda_{1,2}^{SPAC}$ are both purely imaginary). The nature of the \glspl{FP} is intimately related to the shape of the semiclassical energy landscape [Fig.~\ref{fig:2sites}(a,c)], and heavily impacts on the features of the trajectories of the system in the phase space [Fig.~\ref{fig:2sites}(b,d)]. Trajectories starting in the proximity of the \gls{SPAC} will closely orbit around it for $\eta > \eta_c$, and instead drift away from it exponentially rapidly (and eventually come back at later times) if $0 < \eta < \eta_c$. The instability of $\delta = 0$ in the latter case is at the orgin of the \gls{MQST}, that is a average non-zero population imbalance \cite{raghavan1999coherent}. A system initialized in the sourroundings of the \gls{SFC} will instead closely orbit around it for any $\eta > 0$.

Importantly, being the system conservative and being all the \glspl{FP} isolated, linear centers of the dynamics will always be nonlinear centers as well. This powerful information, extendible to $V = 3$ but in stark contrast with $V \ge 4$, guarantees that the solution of the linearized \gls{GPE} is accurate even at long times and for the whole nonlinear \gls{GPE} \eqref{eq: dyn eq GP 2 sites - reduced} when close to a linear center. This is clear from Fig.~\ref{fig:2sites}(c), where we show the energy landscape in the surroundings of the \gls{SPAC} for $\eta>\eta_c$ and one possible trajectory (in blue). Since energy is conserved and the \gls{SPAC} is isolated, the trajectory must necessary be a cycle around the \gls{SPAC}, even at long times.

\subsection{Stability of the uniform configuration ($V \ge 3$) and macroscopic quantum self-trapping}
\label{sec: Stability of the uniform configuration}

We now consider the case of the \gls{UC}, that is we evaluate the Jacobian matrix \eqref{eq: Jacobian, r = 0} for the configuration \eqref{eq: stationarity condition - UC}. This case is particularly relevant since for a \gls{MI} to \gls{SF} quench and $V \gg 1$ the system is initialized in the proximity of the \gls{UC}, so that we observe a \gls{DPT} at $\eta_c^{UC}$.

The non-zero eigenvalues of the Jacobian matrix $J$ read (details in App.~\ref{app: the linearization procedure})
\begin{align}
&\lambda_{+1}^{\pm} = \frac{i\eta \pm \sqrt{4\eta - \eta^2}}{2} \ , \label{eq: eig UC q = +1}\\
&\lambda_{-1}^{\pm} = \frac{-i\eta \pm \sqrt{4\eta - \eta^2}}{2} \ . \label{eq: eig UC q = -1}
\end{align}
For $\eta < \eta_c^{UC} = 4$ some eigenvalues ($\lambda_1^{+}, \lambda_{-1}^{+}$) have positive real part and some others ($\lambda_1^{-}, \lambda_{-1}^{-}$) negative real part, making the \gls{UC} a saddle point of the dynamics. If the system is initialzed in the proximity of the \gls{UC}, it will drift away from it aligning along the direction defined by the two eigenvectors associated to the eigenvalues with positive real part ($\lambda_1^{+}, \lambda_{-1}^{+}$). Along this direction and at short times the modulus $r$ of $\Psi$ grows as (details in App.~\ref{app: the linearization procedure})
\begin{equation}
r \sim e^{t/\tau^{UC}} \ ,
\label{eq: exp growth of r for UC}
\end{equation}
with characteristic timescale (dashed blue line in Fig.~\ref{fig:divergence time}(b))
\begin{equation}
\tau^{UC} = \frac{2}{\sqrt{4\eta - \eta^2}}
\label{eq: characteristic tau UC}
\end{equation}
and where we used the symbol $\sim$ meaning that the exponential divergence will occur after a possible very short transient in which the system aligns with the unstable eigenvector. Such short transient and the exponential growth of $r$ at short times for $\eta < \eta_c^{UC}$ are correctly observed for $\eta = 2$ in the inset of Fig.~\ref{fig:long time and synchronization}(d), with logaritmic ordinate axis. Close to the \gls{DPT} we have $\tau^{UC} \sim \left|1 - \frac{\eta}{\eta_c}\right|^{-\beta}$ with critical exponent $\beta = 1/2$. For $\eta < \eta_c^{UC}$, the increase of $r$ corresponds to an increase of the spread of the boson numbers at each site $\{\rho_j\}$ around their mean value $\rho_0$ [see Sec.~\ref{section: The effects of conserved quantities} and Fig.~\ref{fig:long time and synchronization}(b)], that is to a symmetry breaking and the emergence of \gls{MQST}. In particular, from equation \eqref{eq: consistency conditions bis 1} we got that the variance over the sites of the number of bosons $\{\rho_j\}$ at each site reads $\langle (\rho_j-\rho_0)^2 \rangle_j = 2 \eta (r^2-r_0^2)$ and thus grows as $\sim e^{2t/\tau^{UC}}$ at short times.

For $\eta > \eta_c^{UC} = 4$ instead $\lambda_1^{+}, \lambda_1^{-}, \lambda_{-1}^{+}, \lambda_{-1}^{-}$ are all purely imaginary, the \gls{UC} is a linear center and, at least at short times, the system cycles around it. For a \gls{MI} to \gls{SF} quench and large $V$, $r$ will corrispondingly remain small ($\sim 1/\sqrt{V}$) and fluctuate in time.

\subsection{Stability of the symmetric $\pi$-aligned configuration (even $V$)}
\label{section: stability of the SPAC}

To obtain information on the stability of the \gls{SPAC}, assuming an even number of lattice sites $V$, we diagonalize exactly the Jacobian matrix $J$ evaluated in the configuration \eqref{eq: stationarity condition - SPAC}, finding (see App.~\ref{app: the linearization procedure}) the following non-zero eigenvalues
\begin{align}
&\lambda^{+} = +\sqrt{\eta\left(2 - \eta\right)} \ , \\
&\lambda^{-} = -\sqrt{\eta\left(2 - \eta\right)} \ ,
\end{align}
For a system initialized in the proximity of the \gls{SPAC} the \gls{DPT} is thus located at $\eta_c^{SPAC} = 2$. For $\eta < \eta_c^{SPAC}$ the divergence timescale is $\tau^{SPAC} = 1/\lambda^{+} = \left(\eta(2-\eta)\right)^{-1/2}$ (continuous red line in Fig.~\ref{fig:divergence time} (b)). 

\subsection{Stability of the delta configuration ($V = 4$)}
\begin{figure}[t]
	\begin{center}
		\includegraphics[width=\linewidth]{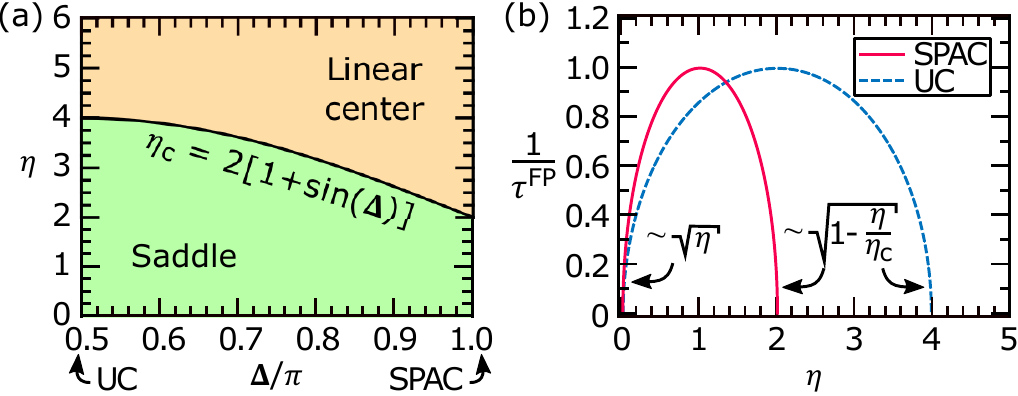}\\
	\end{center}
	\vskip -0.5cm \protect\caption[]
	{(color online) Depending on the value of the dimensionless hopping strength $\eta$, for a number of lattice sites $V \ge 4$, the behavior of a system initialized close to a $r = 0$ \gls{FP} changes sharply in what we refer to as a \gls{DPT}. (a) The \gls{DC} (defined for $V = 4$) is a saddle (linear center) of the dynamics if $\eta < \eta_c^{DC} = 2(1 + \sin(\Delta))$ ($\eta > \eta_c^{DC}$). If and only if $\eta < \eta_c^{UC}$ ($\eta < \eta_c^{SPAC}$), for a system initialized in the proximity of the \gls{UC} (\gls{SPAC}), the dynamical order parameter $r$ will grow exponentially at short times as $r\sim e^{t/\tau^{UC}}$ ($r\sim e^{t/\tau^{SPAC}}$). (b) Inverse of the characteristic time $\tau^{FP}$ for both the \gls{UC} (blue dashed line) and the \gls{SPAC} (red continuous line). At the critical $\eta_c^{UC} = 4$ and $\eta_c^{SPAC} = 2$ the characteristic time of the \gls{UC} and the \gls{SPAC}, respectively, diverges.}
	\label{fig:divergence time}
	\vskip -0.5cm
\end{figure}

We now aim, for $V = 4$, to study the linear stability of the \gls{DC}, that runs parametrically over the whole manifold of $r = 0$ \glspl{FP}, ranging from the \gls{UC} to the \gls{SPAC}. With the help of a symbolic manipulation software we plug the configuration \eqref{eq: DC} into the Jacobian matrix \eqref{eq: Jacobian, r = 0}, exactly finding its associated characteristic polynomial
\begin{equation}
P(\lambda) = \lambda^{4}\left(\lambda^4 + \eta (\eta-2) \lambda^2 + \eta^2 \sin(\Delta)\right) \ ,
\end{equation}
Studying the roots of $P(\lambda)$, that are the eigenvalues of the Jacobian, it is easy to show that \begin{equation}
\eta_c^{DC} = 2\left(1 + \sin(\Delta)\right) \ ,
\label{eq: eta_c in DC}
\end{equation}
such that the \gls{DC} corresponding to a given $\Delta$ is a saddle (linear center) of the dynamics if $\eta < \eta_c^{DC}$ ($\eta > \eta_c^{DC}$), as shown in the dynamical phase diagram of Fig.~\ref{fig:divergence time} (a). In agreement with the previous results, we find that $\eta_c = 2$ for the \gls{SPAC} ($\Delta = \pi$) and that $\eta_c = 4$ for the \gls{UC} ($\Delta = \pi/2$).

As a final remark, we stress that the validity of the present linear stability analysis is  limited to short times only. Indeed, in the long time regime the nonlinearities of the \gls{GPE} \eqref{eq: dyn eq GP - rho theta - compact - no disorder} crucially impact on the system dynamics. In particular, for $V\ge4$ and $\eta > \eta_c^{UC}$, the \gls{UC} is a linear center but not necessarily a nonlinear center, since it is non isolated (it is in fact part of the manifold of the $r = 0$ \glspl{FP}). This means that in the long time and nonlinear regime a system initialized in the proximity of the \gls{UC} (as for the \gls{MI} to \gls{SF} quench for $V \gg 1$) can a priori still drift away from it, even for $\eta > \eta_c^{UC}$. This reasoning is peculiar of the $V \ge 4$ case and at the basis of the possible emergence of the $\pi$-synchronization of the bosonic phases $\{\theta_j\}$ that we address in the next Section.

\section{Long time dynamics and $\pi$-synchronization}
\label{section: long time and pi synchronization}
In this Section we go beyond the linear analysis presented above and  investigate the long time ($t\gg 1/|\lambda_{\pm1}^{\pm}|$) nonlinear dynamics. 
To this purpose we solved numerically the \gls{GPE} \eqref{eq: dyn eq GP} with the \textsc{matlab} built-in adaptive ordinary differential equations solver \textsf{ode45}, for $V\gg1$ and a \gls{MI} to \gls{SF} quench, so that the \gls{DPT} is located at a critical dimensionless hopping strength $\eta_c^{UC} = 4$ and that $r_0 = r(t = 0) \sim \frac{1}{\sqrt{V}} \ll 1$. The \gls{TWA} is then considered averaging the mean-field observables over a large number of simulations (each one with different random initial phases $\{\theta_j\}$) to approximately compute the dynamics of the expectation values at any time $t \ge 0$ according to Eq.~\eqref{eq: exp value}. In this way, we find that for $\eta < 4$ ($\eta > 4$) a macroscopic dynamical order parameter $\langle r \rangle$ ($\langle S \rangle$) relaxes to a finite value, the finiteness of $\langle S \rangle$ underlying $\pi$-synchronization of the phases $\{\theta_j\}$, that is their tendency to split into two groups with phase difference $\pi$. With its intuitive phase-space representation and its analytical results, our fully-connected model is consistent with the idea that in high-dimensional lattices the system asymptotically builds up long-range correlations \cite{dziarmaga2012quench} only at low effective temperatures (that is weak quenches, i.e. small $\eta$).

\subsection{Long time dynamics for $\eta<4$}
In the long time regime, the nonlinearities of the \gls{GPE} \eqref{eq: dyn eq GP - rho theta - compact - no disorder} comes into play. If we look at the dynamics of $r$ for a given initial condition (e.g.~Fig.~\ref{fig:long time and synchronization}(a,b) for $\eta = 2$ at $t = 0$ and $t = 1000$, respectively), we find that, after the initial growth predicted by the linear stability analysis of Sec.~\ref{section: short time, linearization and DPT}, $r$ fluctuates in time around a finite value. When considering the average $\langle r(t)\rangle$ over a large number of simulations [in the spirit of Eq.~\eqref{eq: exp value}], these long time fluctuations vanish, revealing an asymptotic relaxion to a finite value (e.g.~$\langle r\rangle \rightarrow 0.38$ for $\eta = 2$), as showed in Fig.~\ref{fig:long time and synchronization}(d). As explained in Sec.~\ref{section: The effects of conserved quantities}, the finiteness of $\langle r \rangle$ at long times corresponds to a well-defined spread of the $\{\rho_j\}$ around their mean value $\rho_0$ (that is \gls{MQST}), that is displayed at $t = 1000$ for one specific initial condition in Fig.~\ref{fig:long time and synchronization}(b). Notice that the growth of $\langle r \rangle$ does not indicate at all a tendency of the system to reach the \gls{SFC} (for which $r = 1$ and $\rho_j = \rho_0$). In Sec.~\ref{section: The effects of conserved quantities} we have in fact shown this to be forbidden by the presence of conserved quantities in the non-equilibrium regime. The relaxation of the system to the \gls{SFC} will possibly happen on much longer timescales thanks to the interaction with the environment, that goes beyond the interests of our study.

\subsection{Long time dynamics for $\eta>4$}
\label{tau>4 long time dynamics}

\begin{figure}[t]
	\begin{center}
		\includegraphics[width=\linewidth]{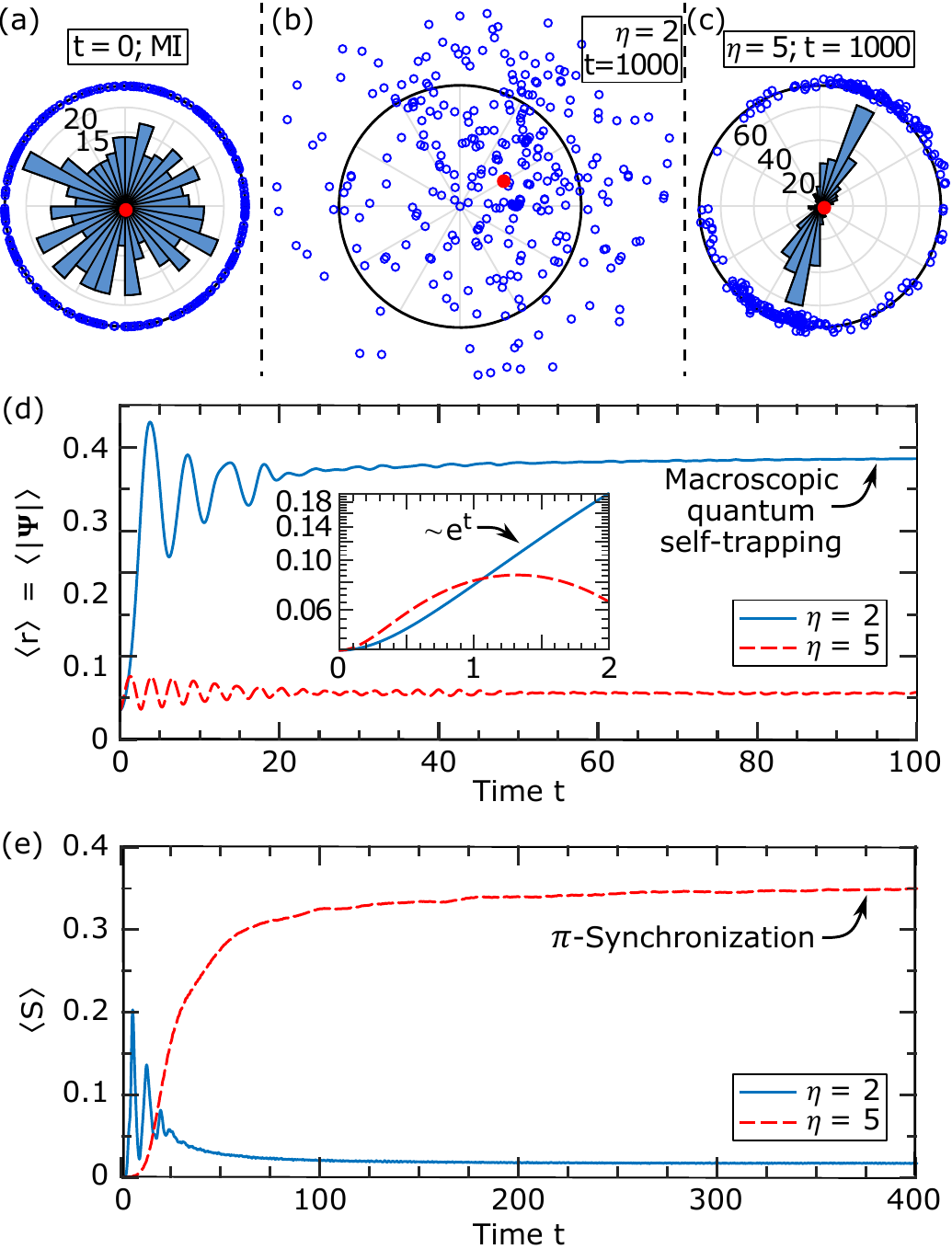}\\
	\end{center}
	\vskip -0.5cm \protect\caption[]
	{(color online)
		Exact numerical solution of the nonlinear dynamics \eqref{eq: dyn eq GP - rho theta - compact - no disorder} up to long times for $V = 500$ sites and a \gls{MI} to \gls{SF} quench.
		(a-c) Graphical representation of the mean-field state for a \emph{single} simulation with initial condition given by Eq.~\eqref{eq: MI}. For graphical clarity, only the blue markers of $300$ out of the $V = 500$ bosonic variables are represented. At time $t = 0$ the phases are randomly distribuited (a) whereas at $t = 1000$ either the $\{\rho_j\}$ are spread around $\rho_0$ (b) or the $\{\theta_j\}$ are $\pi$-synchronized (c).	(d) Dynamics of the expected value $\langle r \rangle$ obtained according to Eq.~\eqref{eq: exp value} as an average over 3000 simulations [each one for a different realization of the random initial phases \eqref{eq: MI}]. For $\eta = 2 < \eta_c^{UC}=4$ (blue continue line), $\langle r \rangle$ grows exponentially at short times ($\langle r\rangle \sim \exp[t/\tau^{UC}]$, see inset with logaritmic ordinate axis) and relaxes to a finite value $0.38$ at long times, indicating \gls{MQST} and reflected in the spread of the $\{\rho_j\}$ in (b). For $\eta > \eta_c^{UC}$ (red dashed line) $\langle r\rangle$ remains instead small. (b) Dynamics of the expected value $\langle S \rangle$ of the $\pi$-synchronization parameter $S$ obtained as an average over 3000 simulations. For $\eta = 5 > \eta_c^{UC}$, $\langle S \rangle$ asymptotically relaxes to a finite value $0.35$, indicating a robust shift of the system towards the \gls{SPAC}.}
	\label{fig:long time and synchronization}
\end{figure}

As shown in Sec.~\ref{section: short time, linearization and DPT} by solving the linearized \gls{GPE}, if $\eta > \eta_c^{UC}$ at short times the system orbits in the phase space around the \gls{UC}, that is in fact a linear center of the dynamics. Correspondingly, $r$ remains small (in the same order of $r_0$, meaning that no \gls{MQST} occurs) and fluctuates, eventually relaxing at long times (dashed red line for $\eta = 5$ in Fig.~\ref{fig:long time and synchronization}(d)). However, in striking contrast with the $V = 2, 3$ cases,  the \gls{UC} is a non-isolated \gls{FP} (it is in fact part of the continuous manifold of $r = 0$ \glspl{FP}), and in general is thus not a nonlinear center of the dynamics, despite the system being conservative. That is, when considering the whole nonlinear \gls{GPE} \eqref{eq: dyn eq GP - rho theta - compact - no disorder}, at long times the system can actually drift away from the initial condition, moving in the proximity of the manifold of the $r = 0$ \glspl{FP} and along it, still conserving energy and total number of particles. For instance, considering the particular initialization at $t = 0$ of Fig.~\ref{fig:long time and synchronization}(a), the system at $t = 1000$ for $\eta = 5$ looks considerably differently but still with $r \ll 1$, [Fig.~\ref{fig:long time and synchronization}(c)]. To track the position of the system in the phase space with respect to the manifold of $r = 0$ \glspl{FP} we introduce therefore a \emph{$\pi$-synchronization} dynamical order parameter $S$ defined as
\begin{equation}
S(t) = \frac{1}{V^2}\sum_{j,k=1}^{V} \psi_k^* \psi_k^* \psi_j \psi_j \ ,
\label{eq: S - dynamical order parameter}
\end{equation}
An interpretation of $S$ is easily accessible expressing it as $S = \frac{1}{V^2}\sum_{j,k=1}^{V} \rho_j\rho_k [2\cos(\theta_j - \theta_k)^2 - 1 ]$. This quantity, which resembles the order parameter typically considered in the study of liquid crystals at equilibrium \cite{prost1995physics},  provides a measure of the tendency of the phases $\{\theta_j\}$ to $\pi$-synchronize. In the sense of Eq.~\eqref{eq: exp value}, averaging $S$ over a large number of different mean-field evolutions we approximate the expectation value of the corresponding quantum operator $\frac{1}{V^2}\sum_{j,k=1}^{V} a_k^\dagger a_k^\dagger a_j a_j$.

On the manifold of the $r = 0$ \glspl{FP}, $S$ ranges from 0 (for the \gls{UC}) to 1 (for the \gls{SPAC}). Looking at the evolution of $S$ we are therefore able to quantify the position of the system with respect to such manifold in time. This can be clearly seen in the case of $V = 4$, for which the parametric \gls{DC} is characterized by $S(\Delta) = \cos^2(\Delta)$, and works analogously for larger $V$. In Fig.~\ref{fig:long time and synchronization}(e) we plot $\langle S \rangle$ against $t$ for $V = 500$. For $\eta = 5 > \eta_c^{UC}$ (red dashed line) $\langle S \rangle$ increases from $0$, corresponding to the initial \gls{MI}, up to a finite value 0.35, underlying the dynamical emergence of $\pi$-synchronization of the bosonic phases $\{\theta_j\}$. Importantly, the growth at short times is not exponential, confirming once more to be intimately connected to the nonlinearities of the \gls{GPE}. Once more, we stress that this phenomenon is enabled by the manifold of isoenergentic $r=0$ \glspl{FP} that, only for $V \ge 4$, opens a channel for the non-equilibrium dynamics connecting the \gls{UC} to the \gls{SPAC}, towards which the system shifts robustly. Notice that, since the system is isolated, the stabilization of $\langle S \rangle$ for large $V$ is an intrinsic property and is not due to the presence of driving and dissipation, as typically considered in the literature \cite{holmes2012synchronization, lee2013quantum, lee2014entanglement, walter2014quantum, bastidas2015quantum, orth2010dynamics, le2018driven, henriet2016quantum}.

\begin{figure}[t]
	\begin{center}
		\includegraphics[width=\linewidth]{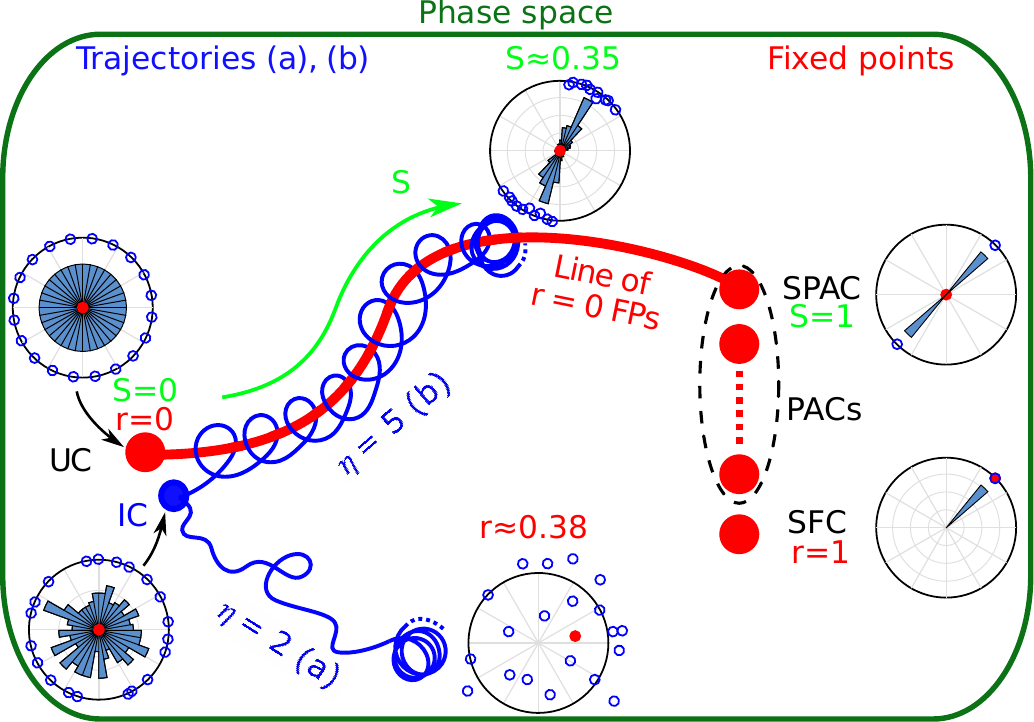}\\
	\end{center}
	\vskip -0.5cm \protect\caption[]
	{(color online) 
		(a,b) Schematic representation of the trajectories of the system (in blue) in the phase space for $\eta = 2$ and $\eta = 5$ respectively, after a \gls{MI} to \gls{SF} quench, under the nonlinear \gls{GPE}, up to long times and for an even $V \gg 1$. The red line and dots represent the \glspl{FP}, and the most relevant configurations are displayed. The blue dot represents the initial condition. For $\eta = 2$ ($\eta = 5$), the system drifts away from (orbits close to) the manifold of the $r = 0$ \glspl{FP}, finally relaxing to a state characterized by a finite $r$ ($S$). For instance, in reference to Fig.~\ref{fig:long time and synchronization}, the state at long time is characterized by $r \approx 0.38$ ($S \approx 0.35$).}
	\label{fig: topology and synchronization}
\end{figure}

In Fig.~\ref{fig: topology and synchronization} we schematically show some possible trajectories of the system (in blue) evolving in the phase space under the \gls{GPE} for a large and even $V$ (the schematic for a odd $V$ would be very similar though, just without the \gls{SPAC}), together with the graphical representation of the most relevant configurations. For a \gls{MI} to \gls{SF} quench and $\eta = 2 < \eta_c^{UC} = 4$ (a), the system drifts away from the manifold of $r = 0$ \glspl{FP} because of the linear instability, eventually relaxing to a state characterized by $\langle r\rangle \approx 0.38$ and by consequent spread of the number of bosons per site $\{\rho_j\}$. Conversely, for $\eta = 5$, the system closely orbits around the aforementioned manifold (since the latter is made of linear centers of the dynamics), while progressively shifting towards the \gls{SPAC} due to the nonlinearities of the \gls{GPE} (b).

\section{Disorder-induced synchronization crossover}
\label{section: competition between hopping strength and disorder}

Having discovered in the previous Section that the dynamics generated by the nonlinear \gls{GPE} can lead at long times to a partial $\pi$-synchronization of the phases $\{\theta_j\}$, we now show that such synchronization does not require fine-tuning of the model parameters, and is rather robust with respect to the introduction of site-dependent disorder. We consider thus the following disordered version of the Bose-Hubbard Hamiltonian
\begin{equation}
H = H_{BH} + \sum_{j=1}^{V} \Omega_j n_j \ ,
\label{eq: disordered BH Hamiltonian}
\end{equation}
where $\{\Omega_j\}$ are a set of independent and identically distributed Gaussian random numbers of zero mean and standard deviation $\Sigma$. Applying a mean-field approximation analogue to the one used to obtain \eqref{eq: dyn eq GP - rho theta - compact - no disorder}, we find the following \gls{GPE} associated to the Hamiltonian \eqref{eq: disordered BH Hamiltonian}
\begin{equation}
\begin{cases}
\displaystyle \frac{d \sqrt{\rho_k}}{dt} = \eta r \sin \left(\theta_k - \phi \right) \\[10pt]
\displaystyle \frac{d \theta_k}{dt} = \eta \frac{r}{\sqrt{\rho_k}}\cos \left(\theta_k -\phi\right) - \rho_k + \omega_k \ ,
\end{cases}
\label{eq: dyn eq GP - rho theta - compact - disorder}
\end{equation}
where $\omega_k = \frac{\Omega_k}{u\rho_0}$. We call disorder strength the dimensionless parameter $\sigma = \frac{\Sigma}{u\rho_0}$, that is the standard deviation of the random numbers $\{\omega_j\}$. Interpreting the variables $\{\theta_j\}$ as the phases of a population of classical oscillators (one per lattice site), in Eq.~\eqref{eq: dyn eq GP - rho theta - compact - disorder} the disorder can be regarded as affecting the oscillators natural frequencies $\{\omega_j\}$, thus competing against the tendency of the oscillators to $\pi$-synchronize. This is reminiscent of the   Kuramoto model for classically coupled nonlinear oscillators \cite{acebron2005kuramoto, kuramoto1975self, strogatz2000kuramoto}.

\begin{figure}[t]
	\begin{center}
		\includegraphics[width=\linewidth]{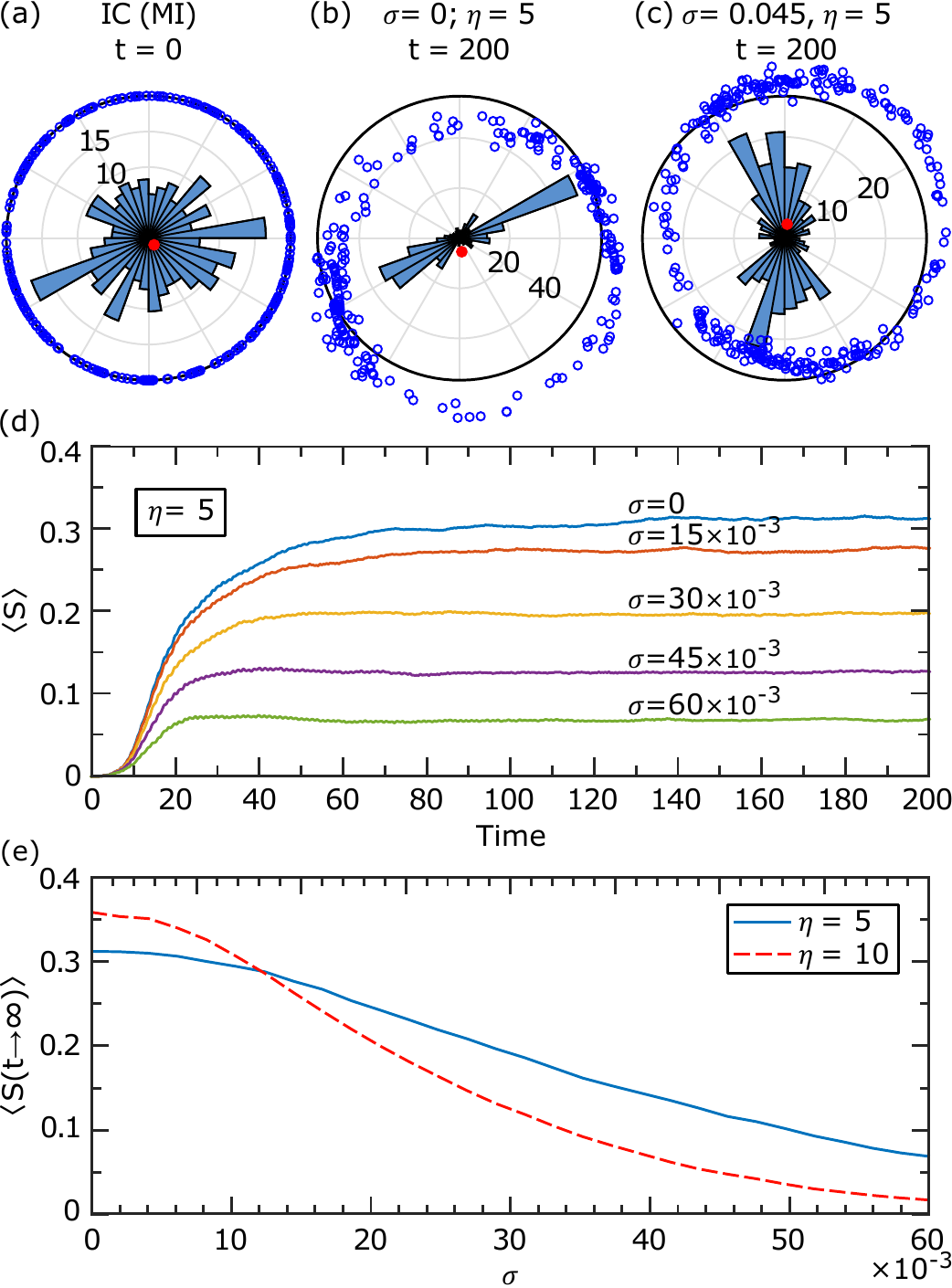}\\
	\end{center}
	\vskip -0.5cm \protect\caption[]
	{(color online) The disorder competes against the tendency of the phases $\{\theta_j\}$ to $\pi$-synchronize. We numerically solve the \gls{GPE} \eqref{eq: dyn eq GP - rho theta - compact - disorder} for $V = 300$ lattice sites and a \gls{MI} to \gls{SF} quench. (a-c) Graphical representation of the mean-field state of the system for a \emph{single} simulation at initial time (a), and at time $t = 200$ for $\sigma=0$ (b) and $\sigma=45\times10^{-3}$ (c). The polar histogram of the phases $\{\theta_j\}$ helps to visualize the reduction of $\pi$-synchronization due to the disorder. (d) Time dynamics of the expected value $\langle S \rangle$ of the $\pi$-synchronization parameter $S$ computed as an average over 1000 simulations [each one with a different realization of the initial random phases \eqref{eq: MI}] according to Eq.~\eqref{eq: exp value}. $\langle S \rangle$ is smaller for larger disorder strengths ($\sigma = 0, 15, 30, 45, 60 \times 10^{-3}$). (e) Asymptotic value $\langle S(t\rightarrow\infty) \rangle$ versus the disorder strength $\sigma$. For increasing disorder, the $\pi$-synchronization is progressively broken in a synchronization crossover with onset decreasing with $\eta$.}
	\label{fig:S vs t disorder}
\end{figure}

To investigate such competition, we solve the \gls{GPE} \eqref{eq: dyn eq GP - rho theta - compact - disorder} for $V = 300$ lattice sites and a \gls{MI} to \gls{SF} quench. Looking at the mean-field dynamics obtained for $\eta = 5$ for one given initial condition [shown in Fig.~\ref{fig:S vs t disorder}(a)], at time $t = 200$ we observe a clear reduction of the $\pi$-synchronization in the disordered case [$\sigma = 0.045$, Fig.~\ref{fig:S vs t disorder}(c)] with respect to the non-disordered one [$\sigma = 0$, Fig.~\ref{fig:S vs t disorder}(a)]. The dynamics of the expectation value $\langle S \rangle$ of the $\pi$-synchronization parameter $S$ is then computed within the \gls{TWA} as an average over the dynamics obtained for 1000 different realization of the initial random phases \eqref{eq: MI}. In Fig.~\ref{fig:S vs t disorder}(d) we show $\langle S \rangle$ to decrease for an increasing disorder strength $\sigma = 0, 15, 30, 45, 60 \times 10^{-3}$. In Fig.~\ref{fig:S vs t disorder}(e) we plot the asymptotic value of $\langle S \rangle$ against the disorder strength for $\eta = 5$ (continue blue line) and $\eta = 10$ (dashed red line), showing a smooth synchronization crossover with onset decreasing with $\eta$. For small (large) disorder $\sigma$, at long time the phases $\{\theta_j\}$ are $\pi$-synchronized (uncorrelated).

\section{Conclusions}
\label{section: conclusions}

In conclusion, we studied the non-equilibrium dynamics induced by a quantum quench to the \gls{SF} regime in the Bose-Hubbard model \eqref{eq: BH Hamiltonian} on a fully-connected (or all-to-all coupled) $V$-dimensional lattice, with potential experimental applications ranging from cold atoms in optical lattices to superconducting qubits. We derived the semiclassical, discrete and nonlinear \gls{GPE} \eqref{eq: dyn eq GP - rho theta - compact - no disorder}, which amounts to a problem of $V$ classical and nonlinearly coupled oscillators with variable phase and length $\{\theta_j, \sqrt{\rho_j}\}$, and accounted for quantum fluctuations considering an ensemble of stochastic initial phases $\{\theta_j\}$ in the so called \gls{TWA}.
Having compacted the \gls{GPE} thanks to the introduction of a complex dynamical order parameter $r$, we showed that for $V\ge4$ there exists a peculiar class of infinitely many \glspl{FP} of the \gls{GPE} (namely the ones with $r = 0$ and homogeneous density $\rho_j = \rho_0$) forming a manifold in the phase space. Among the \glspl{FP} of such manifold, particularly relevant are the \gls{SPAC} \eqref{eq: stationarity condition - SPAC} and the \gls{UC} \eqref{eq: stationarity condition - UC}, the latter being in the proximity of the initial condition in case of a \gls{MI} to \gls{SF} quench for $V\gg1$. Linearizing the \gls{GPE} and diagonalizing the respective Jacobian matrix, we studied the short time dynamics for a system initialized in the proximity of the most relevant \glspl{FP}, that is the \gls{UC} (for $V \ge 3$), the \gls{SPAC} (for even $V$) and the \gls{DC} (a parametric \gls{FP} spanning the $r = 0$ manifold for $V = 4$). We found that, depending on the considered \gls{FP}, there exists a certain critical hopping strength $\eta_c^{FP}$ such that at short times if $\eta > \eta_c^{FP}$ (i.e. small interactions) the system remains close to the initial condition, whereas if $\eta < \eta_c^{FP}$ (i.e. large interactions) it drifts away from the \gls{FP} exponentially fast ($\langle r \rangle\sim e^{t/\tau^{FP}}$), indicating \gls{MQST}. Such sharp change of dynamic behavior when varying $\eta$ across a critical value is a \gls{DPT}, that we located at $\eta_c^{UC} = 4$, $\eta_c^{SPAC} = 2$ and at $\eta_c^{DC} = 2(1 + \sin(\Delta))$. Furthermore, for a \gls{MI} to \gls{SF} quench with $V \gg 1$, we investigated numerically the long time dynamics generated by the whole, nonlinear \gls{GPE}, computing expectation values as averages over the random initial phases according to the \gls{TWA}. For $\eta < \eta_c^{UC} = 4$ we proved the emergence of \gls{MQST} whereas for $\eta > \eta_c^{UC} = 4$, we showed that the system exhibits a  slow drift in the phase space from the proximity of the \gls{UC} towards the \gls{SPAC}, eventually relaxing to a state where the phases $\{\theta_j\}$ are $\pi$-synchronized. We argued this phenomenon, which we quantified with a suitable dynamical order parameter $S$, to be intimately connected to the presence of the manifold of $r = 0$ \glspl{FP}. We finally studied the robustness of the long time $\pi$-synchronization against the introduction in the model of site-dependent disorder, finding that $\langle S \rangle$ vanishes in a smooth synchronization crossover for an increasing disorder strength, meaning that no fine-tuning of the model parameters is needed for the $\pi$-synchronization to occur.

{\it Future developments.}  We conclude by outlining possible developments of the present investigation. A fascinating challenge  is to work out  analytical approaches that enable to understand at a deeper level  the emergence of $\pi$-synchronization, that is intimately connected to the nonlinear terms of the \gls{GPE}. One possibility is represented by the continuum limit for $V \rightarrow \infty$ of the nonlinear \gls{GPE}, that is addressed in App.~\ref{app: V to inf limit and continuous equations}. It is then crucial to analyze the effects beyond the \gls{TWA} due to the finiteness of $\rho_0$ \cite{polkovnikov2003quantum}. Furthermore, we notice that the \gls{TWA} is potentially a powerful tool to address the emergence of spontaneous synchronization in generic (almost) isolated many-bosons quantum systems different  from the one considered in the present work or in Refs. \cite{polkovnikov2002nonequilibrium,dziarmaga2012quench, witthaut2017classical}. As seen, under particular circumstances, such systems can in fact be turned into systems of nonlinearly coupled classical oscillators (in general of variable phase and length), for which synchronization is a universal and fundamental concept \cite{pikovsky2003synchronization}.

\acknowledgments
A.P. acknowledges financial support from the PALM Labex, Paris-Saclay, Grant No.~ANR-10-LABX-0039, and from the Royal Society and the EPSRC. K.L.H acknowledges funding from the ANR BOCA and the Deutsche Forschungsgemeinschaft (DFG, German Research Foundation) via Research Unit FOR 2414 under project No.~277974659. We thank Marco Schiro and Andreas Nunnenkamp for fruitful comments and we also acknowledge discussions at
CIFAR meetings in Canada and at the Centre de Recherches Mathematiques in Montreal.

\appendix

\section{Dynamical equations}
\subsection{Explicitation of the Heisenberg equation of motion}
\label{App: Heiseberg equation}
To compute the commutators of Eq.~\eqref{eq: Heisenberg dynamics} we evaluate the following terms
\begin{equation}
\begin{aligned}
&[n_j(n_j - 1), a_k]
= - 2 n_j a_j\delta_{j,k} \ , \\
&[n_j(n_j - 1), a_k^\dagger]
= + 2 a_j^\dagger n_j \delta_{j,k} \ , \\
&[a_i^\dagger a_j + a_j^\dagger a_i, a_k]
= - \delta_{i,k} a_k - \delta_{j,k} a_k \ , \\
&[a_i^\dagger a_j + a_j^\dagger a_i, a_k^\dagger]
= \delta_{i,k} a_k^\dagger + \delta_{j,k} a_k^\dagger \ ,
\end{aligned}
\end{equation}
so that Eq.~\eqref{eq: Heisenberg dynamics} is explicitly rewritten as
\begin{equation}
\frac{d a_k}{d(it)} = + \frac{J}{V}\sum_{j \neq k}^V a_j - u n_k a_k + \mu a_k \ .
\label{eq: dyn eq 1 - Appendix}
\end{equation}
Exploiting the gauge freedom we can safely operate the following substitution for the bosonic creation and annihilation operators
\begin{equation}
\begin{aligned}
& a_j^\dagger \rightarrow a_j^\dagger e^{- i \Omega_G t} \\
& a_j \rightarrow a_j e^{i \Omega_G t} \ ,
\end{aligned}
\label{eq: change of frame}
\end{equation}
where $\Omega_G$ is an arbitrary real and time-independent number. Under the transformation \eqref{eq: change of frame}, the bosonic commutation relations are in fact preserved, that is $[a_k e^{i \Omega_G t}, a_j^\dagger e^{- i \Omega_G t}] = [a_k, a_j^\dagger] = \delta_{k,j}$, meaning that $a_j e^{i \Omega_G t}$ and $a_j^\dagger e^{- i \Omega_G t}$ are still annihilation and creation bosonic operators associated to the $j$-th site, respectively. Under the gauge transformation \eqref{eq: change of frame}, \eqref{eq: dyn eq 1 - Appendix} transforms into
\begin{equation}
\frac{d a_k}{d(it)} = + \Omega_G a_k + \frac{J}{V} \sum_{j \neq k}^V a_j - u n_k a_k + \mu a_k \ .
\label{eq: dyn eq 2 - Appendix}
\end{equation}
Considering $\Omega_G = \frac{J}{V} - \mu$ we finally get
\begin{equation}
\frac{d a_k}{d(it)} = + \frac{J}{V} \sum_{j=1}^{V} a_j - u n_k a_k \ ,
\end{equation}
that is \eqref{eq: dyn eq 3}.

\subsection{Dynamical equations for $\rho_k$ and $\theta_k$}
\label{App: dynamical equations for rho_k and theta_k}

We can derive the dynamical equations for $\rho_k$ and $\theta_k$, that are the squared modulus and the phase of $\psi_k$, respectively, starting from the ones for $\psi_k$ and $\psi_k^*$ \eqref{eq: dyn eq GP}. We just have to write $\rho_k = \psi_k \psi_k^*$ and $\theta_k = \frac{1}{2i}\log(\frac{\psi_k}{\psi_k^*})$ and proceed with the following straightforward computations
\begin{equation}
\begin{aligned}
\frac{d \rho_k}{d(it)} & = \psi_k^*\frac{d \psi_k}{d(it)} + \psi_k\frac{d \psi_k^*}{d(it)}
= + \frac{J}{V} \sum_{j=1}^{V} \left(\psi_j\psi_k^* - \psi_j^* \psi_k\right) \\
& = + 2 i \frac{J}{V} \sum_{j=1}^{V} \sqrt{\rho_j \rho_k} \sin \left(\theta_j - \theta_k \right) \ ,
\end{aligned}
\end{equation}
\begin{equation}
\begin{aligned}
\frac{d \theta_k}{d(it)} & = \frac{d}{d(it)} \frac{1}{2 i} \log(\frac{\psi_k}{\psi_k^*})
= \frac{1}{2 i} \frac{\frac{\psi_k}{d(it)}\psi_k^* - \frac{\psi_k^*}{d(it)}\psi_k}{|\psi_k|^2} \\
& = - i \left( \frac{J}{V} \sum_{j=1}^{V} \frac{|\psi_j|}{|\psi_k|} \cos \left(\theta_j -\theta_k\right) - u |\psi_k|^2\right) \ , \\
\end{aligned}
\end{equation}
that is Eq.~\eqref{eq: dyn eq GP - rho theta}.

\section{Two-site model}
\label{app: 2sites}
In this Section we review some results on the simple case of $V = 2$ (a bosonic dimer) \cite{milburn1997quantum, smerzi1997quantum, franzosi2000quantum, gati2007bosonic, graefe2008mean, longhi2011optical, albiez2005direct, raghavan1999coherent, chuchem2010quantum}. Exploiting the conservation of the total number of particles $\rho_1 + \rho_2 = 2 \rho_0 = 2$ and introducing the population imbalance $\delta = \rho_1 - \rho_2$ we express the population in the two sites as $\rho_{1,2} = 1 \pm \delta/2$ and reduce the \gls{GPE} \eqref{eq: dyn eq GP - rho theta - compact - no disorder} to a 2-dimensional system of dynamical equations
\begin{equation}
\begin{cases}
\frac{\partial \theta}{\partial t} = - \delta - \frac{\eta}{2} \frac{\delta}{\sqrt{1 - \frac{\delta^2}{4}}} \cos \theta \\[5pt]
\frac{\partial \delta}{\partial t} = 2 \eta \sqrt{1 - \frac{\delta^2}{4}} \sin \theta \ ,
\end{cases}
\label{eq: dyn eq GP 2 sites - reduced.bis}
\end{equation}
that is Eq.~\eqref{eq: dyn eq GP 2 sites - reduced} and where $\theta = \theta_1 - \theta_2$ is the difference of the phases of the two sites. Since $\delta = \pm 2$ corresponds to the case of all particles being in the same site (that is $V = 1$), we consider $|\delta| < 2$ without loss of generality. The stationarity conditions read
\begin{align}
- \delta = \frac{\eta}{2} \frac{\delta}{\sqrt{1 - \frac{\delta^2}{4}}} \cos \theta \ , \label{eq: stationarity condition - 2 sites 1}\\
2 \eta \sqrt{1 - \frac{\delta^2}{4}} \sin \theta = 0 \ . \label{eq: stationarity condition - 2 sites 2}
\end{align}
Of course, the condition \eqref{eq: stationarity condition - 2 sites 1} allows a common rotation of the phases $\theta_1$ and $\theta_2$.

The Eq.~\eqref{eq: stationarity condition - 2 sites 2} is solved either by $\theta = 0$ or by $\theta = \pi$. In the two cases, Eq.~\eqref{eq: stationarity condition - 2 sites 1} reads
\begin{equation}
\begin{aligned}
- \delta = \frac{\eta}{2} \frac{\delta}{\sqrt{1 - \frac{\delta^2}{4}}} &\ \quad \text{for} \ \theta = 0 \ , \\
+ \delta = \frac{\eta}{2} \frac{\delta}{\sqrt{1 - \frac{\delta^2}{4}}} &\ \quad \text{for} \ \theta = \pi \ ,
\end{aligned}
\end{equation}
respectively, reading the following four \glspl{FP}
\begin{equation}
\begin{aligned}
& \theta = 0,   \quad \delta = 0 &\text{SFC} \ \ \ , \\
& \theta = \pi, \quad \delta = 0   &\text{SPAC} \ , \\
& \theta = \pi, \quad \delta = \pm \sqrt{4-\eta^2}   &\text{PAC}_\pm \ .
\end{aligned}
\label{eq: 2 sites SCs}
\end{equation}
Of course PAC$_+$ and PAC$_-$ exist only for $\eta < 2$, since $\delta$ and $\theta$ are real variables. As well, notice that the existence and the nature of the \glspl{FP} generally depends on the sign of the on-site interaction $u$. In the present work we only focus on repulsive in-situ interaction ($u, \eta > 0$).

To study the stability of the \glspl{FP}, we diagonalize the $2\time2$ dimensional Jacobian matrix $J$ associated to Eq. \eqref{eq: dyn eq GP 2 sites - reduced.bis}. For the various \glspl{FP} we find the following Jacobian matrices and associated eigenvalues $\lambda_{1,2}$
\begin{equation}
	J^{SFC} = 
	\begin{pmatrix}
	0 & - \frac{\eta}{2} - 1 \\
	2\eta & 0
	\end{pmatrix}
	\ \rightarrow \
	\lambda_{1,2}^{SFC} = \pm 2 i \sqrt{ \frac{\eta}{2} + \frac{\eta^2}{4}} \ ,
\end{equation}	
\begin{equation}
	J^{SPAC} = 
	\begin{pmatrix}
	0 & \frac{\eta}{2} - 1 \\
	- 2\eta & 0
	\end{pmatrix}
	\ \rightarrow \
	\lambda_{1,2}^{SPAC} = \pm 2 \sqrt{\frac{\eta}{2} - \frac{\eta^2}{4}} \ ,
\end{equation}
\begin{equation}
	J^{PAC_\pm} = 
	\begin{pmatrix}
	0 & -\frac{4}{\eta^2} - 1 \\
	- 2\eta & 0
	\end{pmatrix}
	\ \rightarrow \
	\lambda_{1,2}^{PAC_\pm} = \pm 2 i \sqrt{\frac{\eta}{2} + \frac{2}{\eta}} \ .
\end{equation}
Notice that, as shown in Sec.~\ref{section: stability of the SPAC}, considering the $4\times4$-dimensional Jacobian \eqref{eq: Jacobian.0} would just have generated 2 additional zero eigenvalues, that are nevertheless not relevant for the determination of the stability of the \glspl{FP}.

\section{Linear stability analysis}
\label{app: the linearization procedure}
In this Section we derive the Jacobian matrix associated to the linearized \gls{GPE} and diagonalize it exactly, finding its eigenvalues (and eventually eigenvectors), for any $V$ and for the various relevant \glspl{FP}, that are the \gls{UC}, the \gls{SPAC}, and the \gls{SFC}. Since the stationarity condition \eqref{eq: stationarity condition 2} allows a common rotation of all the phases at rate $\Omega$, we move to a frame rotating exactly at the angular speed $\Omega$ associated to the considered \gls{FP}, where the equations of motion \eqref{eq: dyn eq GP - rho theta - compact - no disorder} read
\begin{equation}
\begin{cases}
\frac{\partial \theta_j}{\partial t} = -x_j^2 +  \eta r \frac{1}{x_j} \cos \left(\phi - \theta_j\right) - \Omega\\
\frac{\partial x_j}{\partial t} = - \eta r \sin \left(\phi - \theta_j\right) \ ,
\label{eq: dyn eq GP - rho theta - compact - no disorder - rotating frame}
\end{cases}
\end{equation}
with $x_j = \sqrt{\rho_j}$. In the new frame we recover the familiar definition of \gls{FP}, reading $\frac{d \theta_k}{d t} = 0$ (i.e. the phases of a \gls{FP} are not rotating). Since the first and the second halves of the state vector $\vec{y}$ refer to the phases and to the moduli of the mean-field bosonic variables, respectively, it is convenient to distinguish the following 4 terms of the Jacobian \eqref{eq: Jacobian.0}
\begin{equation}
\begin{aligned}
J_{j, k} &=  \frac{\partial}{\partial \theta_k} \left(\frac{d \theta_j}{dt}\right) \ ,\\
J_{j+V, k+V} &=  \frac{\partial}{\partial x_k} \left(\frac{d x_j}{dt}\right) \ ,\\
J_{j+V, k} &=  \frac{\partial}{\partial \theta_k} \left(\frac{d x_j}{dt}\right) \ , \\
J_{j, k+V} &=  \frac{\partial}{\partial x_k} \left(\frac{d \theta_j}{dt}\right) \ ,
\end{aligned}
\label{eq: Jacobian in app}
\end{equation}
To build the Jacobian matrix we evaluate the following partial derivatives
\begin{equation}
\begin{aligned}
&\frac{\partial}{\partial \theta_k} r \cos\left(\phi - \theta_j\right) = \frac{-x_k}{V} \sin \left(\theta_k - \theta_j\right) + \delta_{k,j} r \sin \left(\phi -\theta_j \right) ,\\
&\frac{\partial}{\partial \theta_k} r \sin\left(\phi - \theta_j\right) = \frac{x_k}{V} \cos \left(\theta_k - \theta_j\right) - \delta_{k,j}r \cos \left(\phi -\theta_j \right) ,\\
&\frac{\partial}{\partial x_k} r \cos\left(\phi - \theta_j\right) = \frac{1}{V} \cos(\theta_k - \theta_j) \ ,\\
&\frac{\partial}{\partial x_k} r \sin\left(\phi - \theta_j\right) = \frac{1}{V} \sin(\theta_k - \theta_j) \ ,
\end{aligned}
\label{eq: useful derivatives - compact}
\end{equation}
$\delta_{k,j}$ being the Kronecker delta ($\delta_{k,j} = 1$ if $k = j$, $\delta_{k,j} = 0$ else). Using \eqref{eq: useful derivatives - compact}, the Jacobian \eqref{eq: Jacobian in app} reads
\begin{equation}
\begin{aligned}
J_{j, k} &=  \frac{\eta}{V} \left[\frac{x_k}{x_j} \sin \left(\theta_j - \theta_k\right) + \delta_{k,j}\frac{rV}{x_j} \sin \left(\phi -\theta_j \right) \right] ,\\
J_{j+V, k+V} &= -\frac{\eta}{V} \sin(\theta_k - \theta_j) ,\\
J_{j+V, k} &= - \frac{\eta}{V} \left[ x_k \cos \left(\theta_k - \theta_j\right) - \delta_{k,j}r V \cos \left(\phi -\theta_j \right) \right] ,\\
J_{j, k+V} &= - 2\delta_{k,j}x_k + \frac{\eta}{V} \frac{1}{x_j} \cos(\theta_k - \theta_j) + \\
&\quad - \delta_{k,j} \eta r \frac{1}{x_k^2} \cos(\phi - \theta_k) \ .
\end{aligned}
\label{eq: Jacobian}
\end{equation}
In particular, in the case of \glspl{FP} with $r = 0$, like the \gls{UC}, the \gls{SPAC} and the \gls{DC}, \eqref{eq: Jacobian} simplifies to
\begin{equation}
\begin{aligned}
J_{j, k} &=  - \frac{\eta}{V} \sin \left(\theta_k - \theta_j\right) \ , \\
J_{j+V, k+V} &= -\frac{\eta}{V} \sin(\theta_k - \theta_j) \ , \\
J_{j+V, k} &= - \frac{\eta}{V} \cos \left(\theta_k - \theta_j\right) \ , \\
J_{j, k+V} &= + \frac{\eta}{V} \cos(\theta_k - \theta_j) - 2\delta_{k,j} \ ,
\end{aligned}
\end{equation}
that is Eq.~\eqref{eq: Jacobian, r = 0}. Having written explicitly $J$, we now aim to diagonalize it exactly, that is to solve the following eigenvalue problem
\begin{equation}
J\vec{y} = \lambda \vec{y} \ .
\label{eq: eigenvalue problem}
\end{equation}

\subsection{Diagonalization of $J$ for the \gls{UC}}
Since the first and the second halves of the state vector $\vec{y}$ \eqref{eq: state vector} refer to the phases and to the moduli of the mean-field bosonic variables, respectively, it is handful to write $\vec{y}$ as
\begin{equation}
\vec{y} = 
\begin{pmatrix}
\vec{y}^{(1)} \\ \vec{y}^{(2)}
\end{pmatrix}  \ ,
\end{equation}
with $\vec{y}^{(1)}$ and $\vec{y}^{(2)}$ $V$-dimensional column vectors defined by $y_j^{(1)} = \theta_j$ and $y_j^{(2)} = x_j$. Performing the multiplication of the matrix $J$ times the column vector $\vec{y}$ we can thus write
\begin{equation}
\begin{aligned}
\left(J \vec{y}\right)_{j} &= - \frac{\eta}{V} \sum_{k = 1}^{V} \sin(\theta_k - \theta_j)y_k^{(1)} \\
& \quad + \frac{\eta}{V} \sum_{k = 1}^{V} \cos(\theta_k - \theta_j) y_k^{(2)} - 2y_j^{(2)} \ , \\
\left(J \vec{y}\right)_{j+V} &= - \frac{\eta}{V} \sum_{k = 1}^{V} \sin(\theta_k - \theta_j) y_k^{(2)} \\
& \quad - \frac{\eta}{V} \sum_{k = 1}^{V} \cos(\theta_k - \theta_j) y_k^{(1)} \ .
\end{aligned}
\label{eq: Jacobian multiplication - CS}
\end{equation}
Since the sine and the cosine can be written in terms of exponentials and since for the \gls{UC} $\theta_k = \frac{2 \pi}{V}k$, the form of \eqref{eq: Jacobian multiplication - CS} suggests us to introduce the following discrete Fourier transform
\begin{equation}
\tilde{v}_q = \frac{1}{V} \sum_{k = 1}^{V} e^{iq\frac{2\pi}{V}k} v_k \quad q \in \mathbb{Z} \ ,
\label{eq: FT}
\end{equation}
where $\vec{v}$ is a $V$-dimensional vector and where we denoted $q$ the Fourier wavenumber. It is easy to verify that $\tilde{v}_{q_1} = \tilde{v}_{q_2}$ if $\frac{q_1 - q_2}{V} \in \mathbb{Z}$, so that it is possible to restrict, without loss of generality, $q \in \{0, 1, 2, \dots, V - 1\}$ and to refer to $q = V-1$ as to $q = -1$. Looking at \eqref{eq: Jacobian multiplication - CS} we are thus interested in the evaluation of the following terms
\begin{equation}
\begin{aligned}
\frac{1}{V} \sum_{k = 1}^{V} \sin(\theta_k - \theta_j) v_k &= \Im{\tilde{v}_1e^{-i\theta_j}} \ , \\
\frac{1}{V} \sum_{k = 1}^{V} \cos(\theta_k - \theta_j) v_k &= \Re{\tilde{v}_1e^{-i\theta_j}} \ ,
\end{aligned}
\label{eq: sine & cosine FT}
\end{equation}
where $\Re$ and $\Im$ denote the real and the imaginary part, respectively. Having introduced the Fourier transform \eqref{eq: FT} and having evaluated the terms of \eqref{eq: sine & cosine FT}, we can write Eq.~\eqref{eq: Jacobian multiplication - CS} in the following compact form
\begin{equation}
\begin{aligned}
\left(J \vec{y}\right)_{j} & = -\eta \Im{\tilde{y}_1^{(1)}e^{-i\theta_j}} +\\
& \quad + \eta \Re{\tilde{y}_1^{(2)}e^{-i\theta_j}} - 2y_j^{(2)} \ , \\
\left(J \vec{y}\right)_{j+V} & = -\eta \Im{\tilde{y}_1^{(2)}e^{-i\theta_j}} +\\
& \quad -\eta \Re{\tilde{y}_1^{(1)}e^{-i\theta_j}} \ ,
\end{aligned}
\end{equation}
that allows us to write the eigenvalue problem \eqref{eq: eigenvalue problem} as
\begin{equation}
\begin{cases}
\lambda y_{j}^{(1)} = -\eta \Im{\tilde{y}_1^{(1)}e^{-i\theta_j}} + \eta \Re{\tilde{y}_1^{(2)}e^{-i\theta_j}} - 2y_j^{(2)} \\[10pt]
\lambda y_{j}^{(2)} = -\eta \Im{\tilde{y}_1^{(2)}e^{-i\theta_j}} -\eta \Re{\tilde{y}_1^{(1)}e^{-i\theta_j}} \ .
\end{cases}
\label{eq: eigenvalue equation}
\end{equation}
The solution of \eqref{eq: eigenvalue equation} will provide us with the Jacobian eigenvalues $\{\lambda_n\}$. Eq.~\eqref{eq: eigenvalue equation} can be approached performing a Fourier transform on it. To do it, we evaluate the following terms
\begin{equation}
\begin{aligned}
& \left(\Im{A e^{-i\theta_j}}\right)_1  = \frac{1}{V} \sum_{j = 1}^{V} \frac{A - A^* e^{2i\theta_j}}{2i} = -i\frac{A}{2} \ ,
\\
& \left(\Re{A e^{-i\theta_j}}\right)_1  = \frac{1}{V} \sum_{j = 1}^{V} \frac{A + A^* e^{2i\theta_j}}{2} = \frac{A}{2} \ ,
\\
& \left(\Im{A e^{-i\theta_j}}\right)_{-1}  = \frac{1}{V} \sum_{j = 1}^{V} \frac{Ae^{-2i\theta_j} - A^*}{2i} = +i\frac{A^*}{2} \ ,
\\
& \left(\Re{A e^{-i\theta_j}}\right)_{-1}  = \frac{1}{V} \sum_{j = 1}^{V} \frac{Ae^{-2i\theta_j} + A^*}{2} = \frac{A^*}{2} \ ,
\\
&\left(\Im{A e^{-i\theta_j}}\right)_{q}  = \frac{1}{V} \sum_{j = 1}^{V} \frac{Ae^{i(q-1)\theta_j} - A^*e^{i(q+1)\theta_j}}{2i} = 0 \ ,
\\
&\left(\Re{A e^{-i\theta_j}}\right)_{q}  = \frac{1}{V} \sum_{j = 1}^{V} \frac{Ae^{i(q-1)\theta_j} + A^*e^{i(q+1)\theta_j}}{2} = 0 \ ,
\end{aligned}
\label{eq: FT of Im and Re}
\end{equation}
$A$ being an arbitrary complex number and $\left(\bullet_j\right)_q$ being an alternative notation for the Fourier transform of the function $\bullet_j$ with respect to the Fourier wavenumber $q = 0, 1, 2, \dots, V - 1$ (that is $\left(\bullet_j\right)_q = \tilde{\bullet}_q$). Importantly, we notice that expressions \eqref{eq: FT of Im and Re} for $q = \pm 1$ are valid if and only if $V \ge 3$, since $\sum_{j = 1}^{V} Ae^{\pm2i\theta_j} \neq 0$ for $V = 2$. We therefore assume for the following treatment that $V \ge 3$. Performing the Fourier transform of \eqref{eq: eigenvalue equation} for $q = \pm 1$ and exploiting the expressions \eqref{eq: FT of Im and Re}, we get
\begin{align}
&\begin{cases}
\lambda \tilde{y}_1^{(1)} & = i\frac{\eta}{2} \tilde{y}_1^{(1)} + \left(\frac{\eta }{2}-2\right)\tilde{y}_1^{(2)} \\
\lambda \tilde{y}_1^{(2)} &= i\frac{\eta}{2} \tilde{y}_1^{(2)} -\frac{\eta }{2}\tilde{y}_1^{(1)} \ ,
\end{cases} \label{eq: eigenvalue problem q = 1}\\
&\begin{cases}
\lambda \tilde{y}_{-1}^{(1)} & = -i\frac{\eta}{2} \tilde{y}_{-1}^{(1)} + \left(\frac{\eta }{2}-2\right)\tilde{y}_{-1}^{(2)} \\
\lambda \tilde{y}_{-1}^{(2)} &= -i\frac{\eta}{2} \tilde{y}_{-1}^{(2)} -\frac{\eta }{2}\tilde{y}_{-1}^{(1)} \ ,
\end{cases} \label{eq: eigenvalue problem q = -1}
\end{align}
that are 2-dimensional eigenvalue problems for $\tilde{y}_1^{(1)}, \tilde{y}_1^{(2)}$ and for $\tilde{y}_{-1}^{(1)}, \tilde{y}_{-1}^{(2)}$ respectively and where we recall the subscripts $\pm 1$ to refer to the Fourier wavevector $q$ and the superscripts $1,2$ to refer to the bipartition of $\vec{y}$ in it first and second halves. We rewrite the problems \eqref{eq: eigenvalue problem q = 1} and \eqref{eq: eigenvalue problem q = -1} in matricial form as
\begin{align}
&\begin{pmatrix}
i \frac{\eta}{2} & \left(\frac{\eta}{2} - 2\right)  \\
-\frac{\eta}{2} & i\frac{\eta}{2}
\end{pmatrix}
\begin{pmatrix}
\tilde{y}_1^{(1)} \\
\tilde{y}_1^{(2)}
\end{pmatrix}
=\lambda 
\begin{pmatrix}
\tilde{y}_1^{(1)} \\
\tilde{y}_1^{(2)}
\end{pmatrix} \ , \label{eq: eigenvalue equation, q = 1}\\
&\begin{pmatrix}
-i \frac{\eta}{2} & \left(\frac{\eta}{2} - 2\right)  \\
-\frac{\eta}{2} & -i\frac{\eta}{2}
\end{pmatrix}
\begin{pmatrix}
\tilde{y}_{-1}^{(1)} \\
\tilde{y}_{-1}^{(2)}
\end{pmatrix}
=\lambda 
\begin{pmatrix}
\tilde{y}_{-1}^{(1)} \\
\tilde{y}_{-1}^{(2)}
\end{pmatrix}  \ , \label{eq: eigenvalue equation, q = -1}
\end{align}
and find the respective eigenvalues $\lambda_{+ 1}^{\pm}$ and $\lambda_{- 1}^{\pm}$ and eigenvectors $v_{+ 1}^{\pm}$ and $v_{- 1}^{\pm}$

\begin{align}
&\lambda_1^{\pm} = \frac{i\eta \pm \sqrt{4\eta - \eta^2}}{2} \ ,
&\lambda_{-1}^{\pm} = \frac{-i\eta \pm \sqrt{4\eta - \eta^2}}{2}  \ ,\\
&v_1^{\pm} = \left(\mp\sqrt{\eta(4-\eta)},\eta\right)^T \ ,
&v_{-1}^{\pm} = v_1^{\pm} \ . \label{eq: unstable modes}
\end{align}

We proceed looking for other non-zero eigenvalues, that is for $\lambda \notin \{0, \lambda_1^+, \lambda_1^-, \lambda_{-1}^+, \lambda_{-1}^-\}$. Performing the Fourier transform of equation \eqref{eq: eigenvalue equation} for $q\neq \pm 1$ we get
\begin{equation}
\begin{cases}
\lambda \tilde{y}_q^{(1)} & = - 2\tilde{y}_q^{(2)} \\
\lambda \tilde{y}_q^{(2)} &= 0
\end{cases}
\quad \text{for} \quad q = 0, 2, 3, \dots, V-2 \ ,
\label{eq: eigenvalue equation, q}
\end{equation}
that, assuming $\lambda \neq 0$, is solved by $\tilde{y}_q^{(1)} = \tilde{y}_q^{(2)} = 0 \ \forall \ q = 0, 2, 3, \dots, V-2$. We observe that, if $\lambda \notin \{\lambda_1^+, \lambda_1^-, \lambda_{-1}^+, \lambda_{-1}^-\}$, then $\tilde{y}_1^{(1)} = \tilde{y}_1^{(2)} = \tilde{y}_{-1}^{(1)} = \tilde{y}_{-1}^{(2)} = 0$, since Eq.~\eqref{eq: eigenvalue equation, q = 1} and \eqref{eq: eigenvalue equation, q = -1} still need to be satisfied. This implies that $\vec{y} = 0$, being all its Fourier components equal to $0$. Thus, we conclude that the only non-zero eigenvalues are $\lambda_1^+, \lambda_1^-, \lambda_{-1}^+, \lambda_{-1}^-$, and that $\lambda_0 = 0$ is an eigenvalue with algebraic multiplicity $m_a = 2V - 4$.

We are now interested in understanding how $r$ grows for a system that is initialized in the proximity of the \gls{UC} for $\eta < \eta_c^{UC}$. Consider a configuration initialized as
\begin{equation}
\begin{aligned}
& \theta_j = \frac{2 \pi}{V} j + \delta_{\theta, j} \ ,\\
& x_j = 1 + \delta_{x, j} \ ,
\end{aligned}
\end{equation}
with $\delta_{\theta, j}, \delta_{x, j} \ll 1$. For such configuration we can write $re^{i\phi}$ as
\begin{equation}
re^{i\phi} = \frac{1}{V} \sum_{j = 1}^{V} x_j e^{i\theta_j} = \frac{1}{V} \sum_{j = 1}^{V}(1 + \delta_{x, j}) e^{i\delta_{\theta, j}} e^{i\frac{2\pi}{V}j} \ ,
\end{equation}
that corresponds to a Fourier transform of the term $(1 + \delta_{x, j})e^{i\delta_{\theta, j}}$. Approximating the exponential at linear order we obtain
\begin{equation}
re^{i\phi} \approx \tilde{\delta}_{x 1} + i\tilde{\delta}_{\theta 1} \ ,
\end{equation}
that is $r$ and $\phi$ can be written in terms of the unstable Fourier modes [Eq.~\eqref{eq: unstable modes}], so that it is easy to conclude that for $\eta < \eta_c = 4$
\begin{equation}
re^{i\phi} \sim \left(-i\sqrt{\eta(4 - \eta)} + \eta\right) e^{\frac{\sqrt{4\eta - \eta^2}}{2} t} \ .
\end{equation}

\subsection{Diagonalization of $J$ for the \gls{SPAC}}

Considering an even $V$ and plugging the configuration \eqref{eq: stationarity condition - SPAC} into Eq.~\eqref{eq: Jacobian, r = 0} we find the following Jacobian matrix for the \gls{SPAC}
\begin{equation}
\begin{aligned}
J_{j, k} &=  0 \ , \\
J_{j+V, k+V} &= 0 \ , \\
J_{j+V, k} &= - \frac{\eta}{V} \nu_k \nu_j \ , \\
J_{j, k+V} &= + \frac{\eta}{V} \nu_k \nu_j - 2\delta_{k,j} \ ,
\end{aligned}
\label{eq: Jacobian, SPAC}
\end{equation}
where, after a proper permutation of the sites, $\nu_k = 1$ for $k = 1, \dots, V/2$ and $\nu_k = -1$ for $k = V/2 + 1, \dots, V$. It is therefore handy to view $J$ as composed of $V/2 \times V/2$-dimensional blocks and to write a $2V$-dimensional column vector $\vec{y}$ as
\begin{equation}
\vec{y} = \begin{pmatrix}
\vec{y}^{(1)} \\ \vec{y}^{(2)} \\ \vec{y}^{(3)} \\ \vec{y}^{(4)}
\end{pmatrix} \ ,
\end{equation}
$\vec{y}^{(i)}$ being a $V/2$-dimensional column vector. The eigenvalue problem \eqref{eq: eigenvalue problem} reads then
\begin{equation}
J \vec{y} = \begin{pmatrix}
+\frac{\eta}{2} \left(\tilde{y}^{(3)}_0 - \tilde{y}^{(4)}_0\right) - 2 \vec{y}^{(3)} \\
-\frac{\eta}{2} \left(\tilde{y}^{(3)}_0 - \tilde{y}^{(4)}_0\right) - 2 \vec{y}^{(4)} \\
+\frac{\eta}{2} \left(-\tilde{y}^{(1)}_0 + \tilde{y}^{(2)}_0\right) \\
-\frac{\eta}{2} \left(-\tilde{y}^{(1)}_0 + \tilde{y}^{(2)}_0\right) \\
\end{pmatrix}
= \lambda
\begin{pmatrix}
\vec{y}^{(1)} \\
\vec{y}^{(2)} \\
\vec{y}^{(3)} \\
\vec{y}^{(4)} \\
\end{pmatrix} \ ,
\label{eq: eigenvalue equation for PA}
\end{equation}
where we introduced $\tilde{y}^{(i)}_0 = \frac{2}{V} \sum_{j = 1}^{V/2} \vec{y}^{(i)}_j$.

Looking for non-zero eigenvalues, that is for $\lambda \neq 0$, we readily obtain $\vec{y}^{(2)} = - \vec{y}^{(1)}$ and $\vec{y}^{(4)} = - \vec{y}^{(3)}$, reducing the problem to
\begin{equation}
\begin{cases}
+\eta \tilde{y}^{(3)}_0 - 2 \vec{y}^{(3)} &= \lambda \vec{y}^{(1)}\\
-\eta \tilde{y}^{(1)}_0 &= \lambda \vec{y}^{(3)} \ .
\label{eq: eigenvalue problem SPAC.2}
\end{cases}
\end{equation}
The second equation of \eqref{eq: eigenvalue problem SPAC.2} implies the components of $\vec{y}^{(3)}$ to be all equal, that is $y^{(3)}_j = \tilde{y}^{(3)}_0 \ \forall \ j = 1, 2, \dots, V/2$ and $\eta \tilde{y}^{(1)}_0 = -\lambda \tilde{y}^{(3)}_0$. From the first equation of \eqref{eq: eigenvalue problem SPAC.2} we get that also all the components of $\vec{y}^{(1)}$ are equal, and we are thus left with
\begin{equation}
-\eta \left(\eta - 2\right) \tilde{y}^{(1)}_0 = \lambda^2 \tilde{y}^{(1)}_0 \ .
\end{equation}
Since we look for non-trivial solutions (that is with non-zero $\vec{y}$), we consider $\tilde{y}^{(1)}_0 \neq 0$ and finally obtain the eigenvalues
\begin{equation}
\lambda^{\pm} = \pm \sqrt{\eta\left(2 - \eta\right)} \ .
\end{equation}
Thus, we conclude that the only non-zero eigenvalues are $\lambda^+$ and $\lambda^-$, and that $\lambda_0 = 0$ is an eigenvalue with algebraic multiplicity $m_a = 2V - 4$.

\subsection{Diagonalization of $J$ for the \gls{SFC}}

We now study the stability of the \gls{SFC}. The argument of Sec.~\ref{section: The effects of conserved quantities} is actually sufficient to state that the \gls{SFC} is a nonlinear center of the dynamics for any $\eta>0$, but for completeness we report here a direct and instructive study of its stability by means of the diagonalization of its Jacobian matrix. For the \gls{SF} \eqref{eq: stationarity condition - SPAC}, the Jacobian \eqref{eq: Jacobian} reads
\begin{equation}
\begin{aligned}
J_{j, k} &=  0 \ ,\\
J_{j+V, k+V} &= 0 \ ,\\
J_{j+V, k} &= - \frac{\eta}{V} + \eta \delta_{k,j} \ ,\\
J_{j, k+V} &= + \frac{\eta}{V} - (2 + \eta) \delta_{k,j} \ ,
\end{aligned}
\label{eq: Jacobian, AC}
\end{equation}
It is therefore again natural to write a $2V$-dimensional column vector $\vec{y}$ as $\vec{y} = \begin{pmatrix} \vec{y}^{(1)} \\ \vec{y}^{(2)} \end{pmatrix}$, $\vec{y}^{(i)}$ being a $V$-dimensional column vector. The eigenvalue problem \eqref{eq: eigenvalue problem} is rewritten as
\begin{equation}
J \vec{y} = \begin{pmatrix}
+\eta \tilde{y}^{(2)}_0 - (2 + \eta) \vec{y}^{(2)} \\
-\eta \tilde{y}^{(1)}_0 + \eta \vec{y}^{(1)}
\end{pmatrix}
= \lambda
\begin{pmatrix}
\vec{y}^{(1)} \\
\vec{y}^{(2)} \\
\end{pmatrix} \ ,
\label{eq: eigenvalue equation for AC}
\end{equation}
where $\tilde{y}^{(i)}_0 = \frac{1}{V} \sum_{j = 1}^{V} \vec{y}^{(i)}_j$. Looking for non-zero eigenvalues, that is $\lambda \neq 0$, we can multiply the first equation of \eqref{eq: eigenvalue equation for AC} by $\lambda$, getting
\begin{equation}
\begin{cases}
+\eta \lambda \tilde{y}^{(2)}_0 - (2 + \eta) \lambda \vec{y}^{(2)} = \lambda^2 \vec{y}^{(1)} \\
-\eta \tilde{y}^{(1)}_0 + \eta \vec{y}^{(1)} = \lambda \vec{y}^{(2)} \ .
\end{cases}
\label{eq: eigenvalue equation for AC.1}
\end{equation}
Plugging $\lambda \tilde{y}^{(2)}_0$ from the second equation of \eqref{eq: eigenvalue equation for AC.1} into the first one, we get
\begin{equation}
- (2 + \eta) \left(-\eta \tilde{y}^{(1)}_0 + \eta \vec{y}^{(1)}\right) = \lambda^2 \vec{y}^{(1)} \ ,	
\end{equation}
from which we find $\tilde{y}^{(1)}_0 = 0$, so that
\begin{equation}
- \eta  (2 + \eta) \vec{y}^{(1)} = \lambda^2 \vec{y}^{(1)} \ ,
\end{equation}
giving $\lambda_{1,2} = \pm i \sqrt{\eta(2 + \eta)}$. These are the only non-zero eigenvalues and can therefore be used to argue on the stability of the \gls{SFC}. Since for any value of $\eta>0$ the non-zero eigenvalues are purely imaginary complex conjugate numbers, the \gls{SFC} is a linear center of the dynamics for any $\eta>0$. As we already noticed in Sec.~\ref{section: The effects of conserved quantities}, the \gls{SFC} is actually not only a linear center, but a nonlinear center as well.

\section{$V\rightarrow \infty$ limit and continuous equations}
\label{app: V to inf limit and continuous equations}
We consider the instructive $V\rightarrow\infty$ limit, with potential application in the analytical approach of synchronization phenomena for the phases $\{\theta_j\}$. We replace the discrete site index $j = 1, 2, \dots, V$ with a continuous variable $s \in (0, 2\pi)$, so that the \gls{GPE} \eqref{eq: dyn eq GP - rho theta - compact - no disorder} transform into
\begin{equation}
\begin{cases}
\frac{\partial \sqrt{\rho(s,t)}}{\partial t} = \eta r \sin \left(\theta(s, t) - \phi \right) \\
\frac{\partial \theta(s, t)}{\partial t} = \frac{\eta r}{\sqrt{\rho(s, t)}}\cos \left(\theta(s, t) -\phi\right) - \rho(s, t) \ ,
\end{cases}
\label{eq: dyn eq GP - rho theta - compact - no disorder - continuum - 1}
\end{equation}
where $\Psi = re^{i\phi}$ is redefined as
\begin{equation}
re^{i\phi} = \frac{1}{V} \sum_{j = 1}^{V} \sqrt{\rho_j}e^{i\theta_j} \xrightarrow{V\rightarrow\infty} \frac{1}{2\pi} \int_{0}^{2\pi} ds \sqrt{\rho(s)}e^{i\theta(s)} \ .
\end{equation}
In this way we passed from a system of $2V$ ordinary differential equations in the $2V$ variables $\{\rho_j,\theta_j\}$, to a system of 2 integro-differential equations in the variables $\rho(s, t)$ and $\theta(s, t)$. Notice that for equation \eqref{eq: dyn eq GP - rho theta - compact - no disorder - continuum - 1} to be valid we require as assumption that there exists a permutation of the sites indexes such that the functions $\rho(s,t)$ and $\theta(s,t)$ are continuous, that is such that $\rho_j \xrightarrow{V\rightarrow\infty} \rho_{j+1}$ and $\theta_j \xrightarrow{V\rightarrow\infty} \theta_{j+1} \ \forall \ j = 1, \dots, V$ and $\rho_V \xrightarrow{V\rightarrow\infty} \rho_{1}$ and $\theta_V \xrightarrow{V\rightarrow\infty} \theta_{1}$. This requirement is for instance fulfilled for the initial condition \eqref{eq: MI} of the \gls{MI} to \gls{SF} quench, on which we focus here. In the $V \rightarrow \infty$ limit, the \gls{UC} is defined by
\begin{equation}
\begin{cases}
\theta_{UC}(s, t) = s + \Omega t\\
\rho_{UC}(s, t) = 1 \ ,
\end{cases}
\end{equation}
that, having $r = 0$ and for $\Omega = -1$, is obviously a \gls{FP} of the dynamical equations \eqref{eq: dyn eq GP - rho theta - compact - no disorder - continuum - 1}. We move to the frame rotating at angular frequency $\Omega$ and express the state of the system as
\begin{equation}
\begin{cases}
\theta(s) = s + \xi(s) \\
\sqrt{\rho(s)} = 1 + \delta(s) \ .
\end{cases}
\end{equation}
Importantly, small $\delta$ and $\xi$ correspond to a system being in the proximity of the \gls{UC}, but we do not need to assume it. The equations of motion \eqref{eq: dyn eq GP - rho theta - compact - no disorder - continuum - 1} read
\begin{equation}
\begin{cases}
\frac{\partial \delta(s, t)}{\partial t} &= \eta r \sin \left(\theta(s, t) - \phi \right) \\
\frac{\partial \xi(s, t)}{\partial t} &= \frac{\eta r}{1 + \delta(s, t)}\cos \left(\theta(s, t) -\phi\right) - 2\delta(s, t) - \delta(s, t)^2 \ ,
\end{cases}
\label{eq: dyn eq GP - rho theta - compact - no disorder - continuum - 2}
\end{equation}
where $\Psi = re^{i\phi}$ can be expressed as $re^{i\phi} = \frac{1}{2 \pi}\int_{0}^{2\pi} (1 + \delta) e^{i\xi} e^{is} = \left((1 + \delta) e^{i\xi}\right)_1$, where we denoted $[A(s)]_q = \frac{1}{2 \pi}\int_{0}^{2\pi} A(s) e^{iqs} ds$. That is, we expressed $\Psi$ as the Fourier transform of a composition of the functions $\delta(s,t)$ and $\xi(s,t)$ with respect to the variable $s$. Thus, \eqref{eq: dyn eq GP - rho theta - compact - no disorder - continuum - 2} reads
\begin{equation}
\begin{cases}
\frac{\partial \delta}{\partial t} &= -\eta \Im{\left((1 + \delta) e^{i\xi}\right)_1 e^{-i\theta}} \\
\frac{\partial \xi}{\partial t} &= \frac{\eta}{1 + \delta}\Re{\left((1 + \delta) e^{i\xi}\right)_1 e^{-i\theta}} - 2\delta - \delta^2 \ .
\end{cases}
\label{eq: dyn eq GP - rho theta - compact - no disorder - continuum - 3}
\end{equation}
We Fourier transform the first equation of \eqref{eq: dyn eq GP - rho theta - compact - no disorder - continuum - 3} getting
\begin{equation}
\begin{cases}
\frac{\partial \delta_1}{\partial t} &= i \frac{\eta }{2}\left((1 + \delta) e^{i\xi}\right)_1 \\
\frac{\partial \delta_{-1}}{\partial t} &= - i \frac{\eta }{2}\left((1 + \delta) e^{i\xi}\right)_{-1} \\
\frac{\partial \delta_{q}}{\partial t} &= 0 \quad \forall \ q \in \{0, 2, 3,\dots, V-2\}\\
\frac{\partial \xi}{\partial t} &= \frac{\eta}{1 + \delta}\Re{\left((1 + \delta) e^{i\xi}\right)_1 e^{-i\theta}} - 2\delta - \delta^2 \ .
\end{cases}
\label{eq: dyn eq GP - rho theta - compact - no disorder - continuum - 4}
\end{equation}
Importantly, to go from Eq.~\eqref{eq: dyn eq GP - rho theta - compact - no disorder - continuum - 1} to Eq.~\eqref{eq: dyn eq GP - rho theta - compact - no disorder - continuum - 4} we have introduced no approximations, that is \eqref{eq: dyn eq GP - rho theta - compact - no disorder - continuum - 4} coicides exactly with the \gls{GPE} \eqref{eq: dyn eq GP - rho theta - compact - no disorder}. The form of \eqref{eq: dyn eq GP - rho theta - compact - no disorder - continuum - 4} is particularly convenient since for $q \neq \pm 1$ we find $\delta_q = cst$ (even for the whole nonlinear dynamics). From \eqref{eq: dyn eq GP - rho theta - compact - no disorder - continuum - 4} it is of course possible to study the linear stability of the \gls{UC} considering small $\delta$ and $\xi$, obtaining the Jacobian eigenvalues $\lambda_{+ 1}^+, \lambda_{+ 1}^-, \lambda_{- 1}^+, \lambda_{- 1}^-$ and highlighting a \gls{DPT} at the critical hopping trength $\eta_c^{UC} = 4$ (that is not surprising at all since the results of Sec.~\ref{section: short time, linearization and DPT} are valid for any $V\ge3$). However, the nonlinearities of Eq.~\eqref{eq: dyn eq GP - rho theta - compact - no disorder - continuum - 4} are the fundamental ingredient to try to capture the emergence of $\pi$-syncrhonization for a \gls{MI} to \gls{SF} quench. Such synchronization is encapsulated into the increase of $S$ up to a finite value, that corresponds to the growth of the Fourier components of $\xi$ with even wavenumber $q$. Additionally, a potentially useful idea in analogy with the Kuramoto model \cite{acebron2005kuramoto} is treating $r$ as a parameter and considering a function $p(\theta, t)$ describing the density of oscillators at the angle $\theta$ at time $t$ (a partial differential equation describing the dynamics of $p(\theta,t)$ would then be the continuity equation).

\bibliography{FCBHM_Bibliography}

\end{document}